\documentclass[%
superscriptaddress,
nobibnotes,
amsmath,amssymb,
aps,
prl,
twocolumn
]{revtex4-1}
\usepackage{graphicx}
\usepackage{dcolumn}
\usepackage{bm}

\begin {document}

\title{Wetting hysteresis induces effective unidirectional water transport through a fluctuating nanochannel}

\author{Noriyoshi Arai}
\email{arai@mech.keio.ac.jp}
\affiliation{%
Department of Mechanical Engineering, Keio University, Yokohama, Kanagawa 223-8522, Japan
}%
\affiliation{%
Computational Astrophysics Laboratory, RIKEN, 2-1 Hirosawa, Wako, Saitama 351-0198, Japan
}%

\author{Eiji Yamamoto}
\affiliation{%
Department of System Design Engineering, Keio University, Yokohama, Kanagawa 223-8522, Japan
}%

\author{Takahiro Koishi}
\affiliation{%
Department of Applied Physics, University of Fukui, 3-9-1 Bunkyo, Fukui 910-8507, Japan
}%

\author{Yoshinori Hirano}
\affiliation{%
Department of Mechanical Engineering, Keio University, Yokohama, Kanagawa 223-8522, Japan
}%

\author{Kenji Yasuoka}
\affiliation{%
Department of Mechanical Engineering, Keio University, Yokohama, Kanagawa 223-8522, Japan
}%

\author{Toshikazu Ebisuzaki}
\affiliation{%
Computational Astrophysics Laboratory, RIKEN, 2-1 Hirosawa, Wako, Saitama 351-0198, Japan
}%



\begin{abstract}
We propose a water pump that actively transports water molecules through nanochannels.
Spatially asymmetric thermal fluctuations imposed on the channel radius cause unidirectional water flow without osmotic pressure, which can be attributed to hysteresis in the cyclic transition between the wetting/drying states.
We show that the water transport depends on fluctuations, such as white, Brownian, and pink noises.
Because of the high-frequency components in white noise, fast switching of open and close states inhibits channel wetting.
Conversely, pink and Brownian noises generate high-pass filtered net flow.
Brownian fluctuation leads to a faster water transport rate, whereas pink noise has a higher capability to overcome osmotic pressure in the opposite direction.
A trade-off relationship exists between the resonant frequency of the fluctuation and the flow amplification.
The proposed pump can be considered as an analogy for the reversed Carnot cycle, which is the upper limit on the energy conversion efficiency.
\end{abstract}

\maketitle

\begin{figure*}
\begin{center}
\includegraphics[width=150 mm,bb= 0 0 1881 1087]{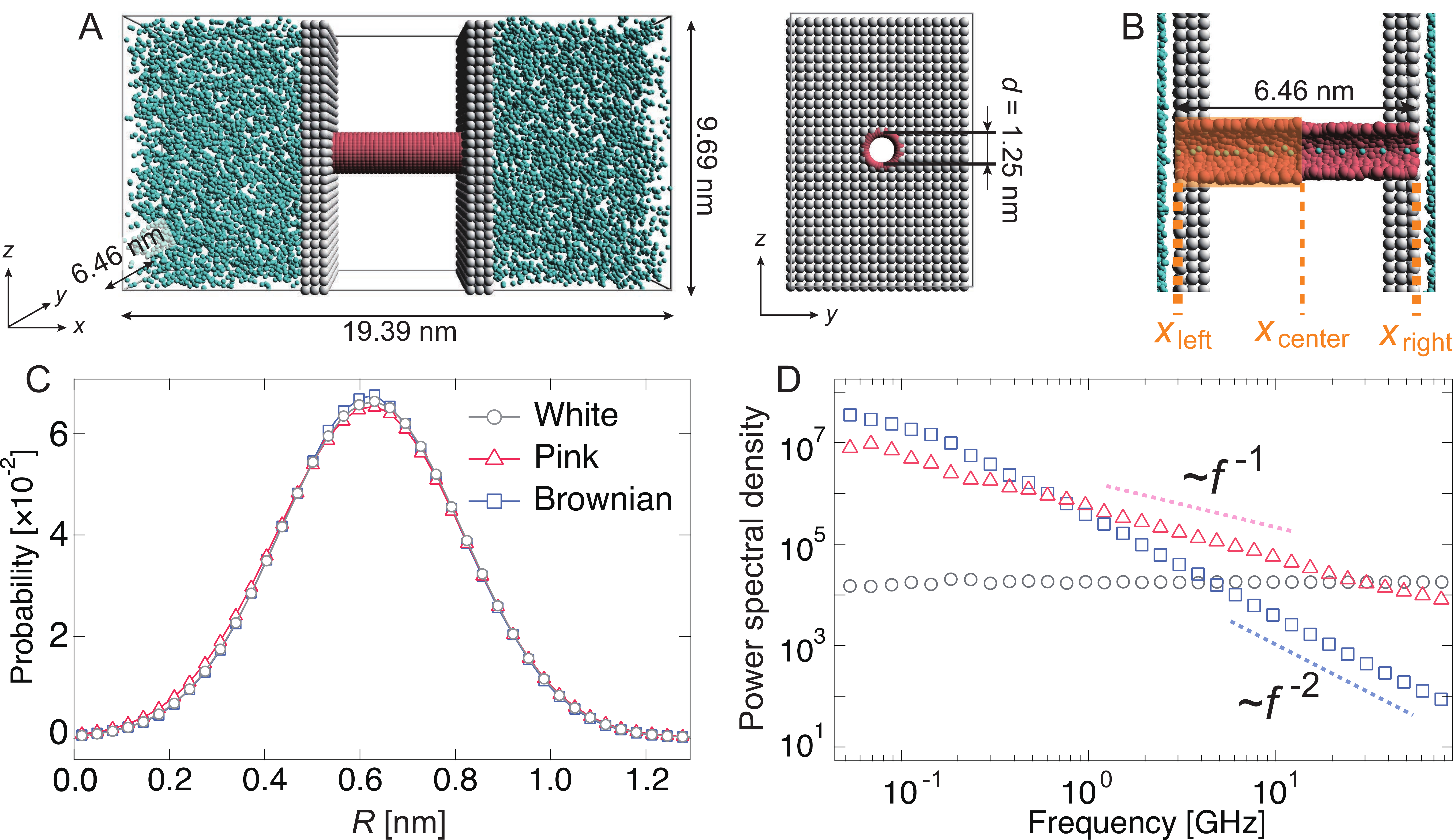}
\caption{Active water pump model.
(A)~Active pump model.
The water reservoirs (green) are separated by a partition (gray) that has a nanochannel (red).
(B)~Cross-sectional view through the center of the nanochannel.
The radius of the left half of the channel (colored orange) changes its size according to certain noise.
The left-end, right-end, and center positions of the nanochannel are labeled as $x_\mathrm{left}$, $x_\mathrm{right}$, and $x_\mathrm{center}$, respectively.
(C)~Probability distributions and (D)~power spectral density (PSD) of the radius size of the left half of the channel.
The trajectory for 80~ns was used for the PSD.
Gray, pink, and blue symbols represent the data obtained using white Gaussian noise, $1/f$ (pink) noise, and $1/f^2$ (Brownian) noise, respectively.}
\label{fig01}
\end{center}
\end{figure*}

\begin{figure*}
\begin{center}
\includegraphics[width=150 mm,bb= 0 0 1323 1162]{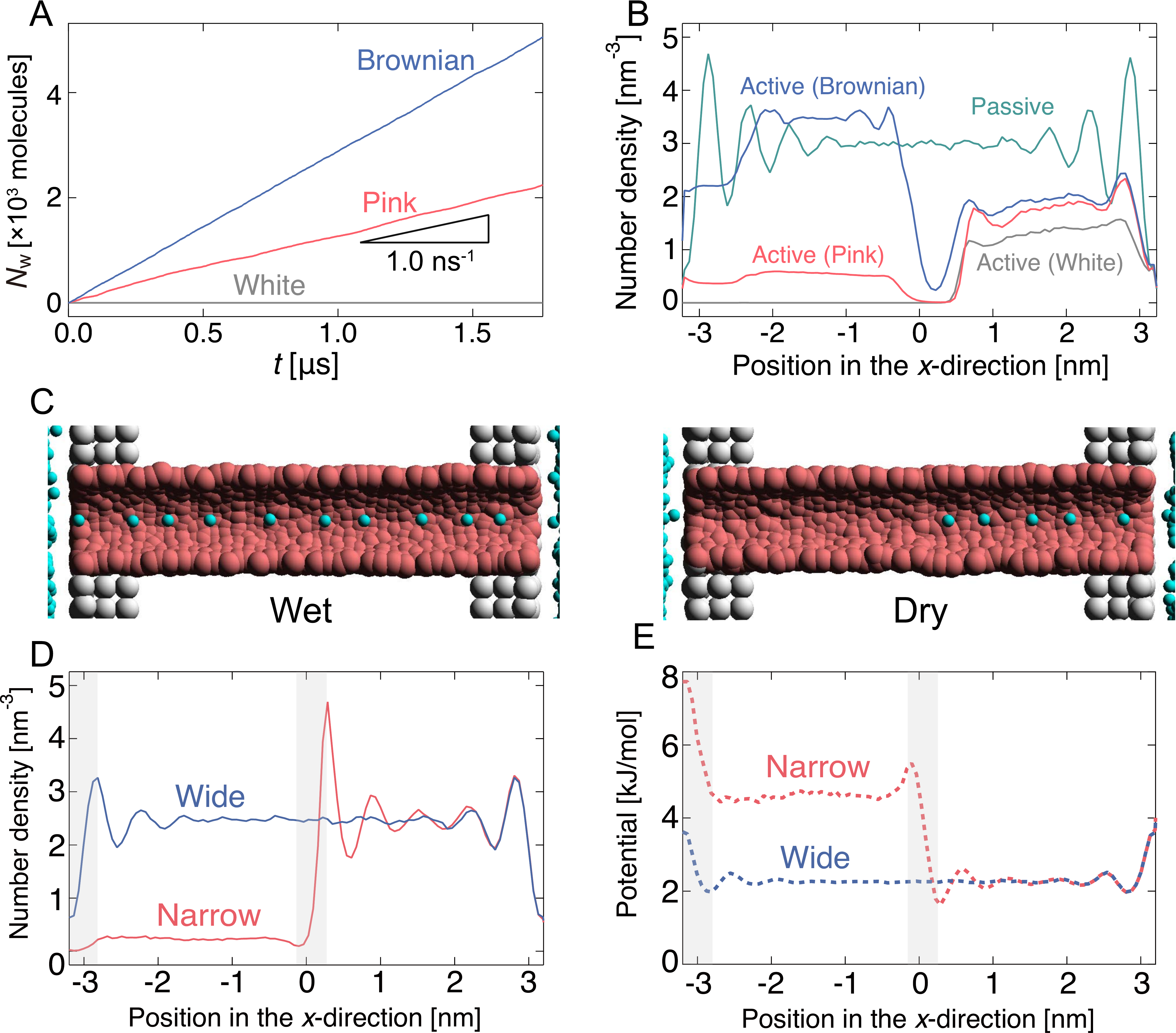}
\caption{Water transport through an active pump.
(A)~Cumulative water transport through a pump driven by thermal fluctuations imposed on the channel radius.
(B)~Number density profiles of the water molecules for passive (green) and active (gray, pink, and blue) states in the $x$-direction.
Here, $x_\mathrm{center}$ is set to zero.
Gray, pink, and blue colored lines represent data obtained using white, $1/f$ (pink), and $1/f^2$ (Brownian) noises, respectively.
Each curve is plotted based on the data averaged over 20 independent simulations.
(C)~Snapshots of wet and dry states in the channel. 
(D)~Water density distribution $\rho(x)$ along the central axis of the channel at equilibrium states.
Blue and red colored curves represent $\rho(x)$ at the wide ($R = 0.623$~nm) and narrow ($R = 0.610$ nm) radii of the channel, respectively.
(E)~Effective potential curves estimated by the water density distributions in the $x$-direction, $\phi(x)  = -k_\mathrm{B}T \mathrm{ln}[\rho(x)/\rho_0]$ where $\rho_0$ is density in bulk.}
\label{fig02}
\end{center}
\end{figure*}


Membranes that can arbitrarily control the amount of water transported must be developed for various applications such as water purification, bio-sensing, and energy harvesting~\cite{WerberOsujiElimelech2016, ShenLiuHanJin2021}.
In particular, various membrane-based or channel-based systems have been used to study water purification and desalination ~\cite{BarboiuGilles2013, HuoZeng2016, LynchRaoSansom2020}.
Aquaporins (AQPs) are membrane proteins that can permeate a few water molecules per nanosecond for a single pore and reject other small molecules~\cite{MurataMitsuokaHiraiWalzAgreHeymannEngelFujiyoshi2000, AgreKingYasuiGugginoOttersenFujiyoshiEngelNielsen2002}.
Hydrophobic channel linings with a well-defined structure can achieve water purification with high permeation efficiency and selectivity in response to osmotic pressure~\cite{GrootGrubmuller2001, SavageOConnellMierckeFiner-MooreStroud2010, GravelleJolyDetcheverryYbertCottin-BizonneBocquet2013, ErikssonFischerFriemannEnkaviTajkhorshidNeutze2013, HornerZocherPreinerOllingerSiliganAkimovPohl2015}.
Mimicry of biological channel proteins, including AQPs and ion channels, inspire novel implementations of artificial water channels~\cite{ShenSiErbakanDeckerDeSaboeKangMajdButlerWalzAksimentievHouKumar2015, LicsandruKocsisShenMurailLegrandVanDerLeeTsaiBaadenKumarBarboiu2016, ShenSongBardenRenLangFerozHendersonSaboeTsaiYanButlerBazanPhillipHickeyCremerVashisthKumar2018, SongJoshiChowdhuryNajemShenLangHendersonTuFarellPitzMaranasCremerHickeySarlesHouAksimentievKumar2020, ShenYeRomaniesRoyChenRenLiuZeng2020, PorterWerberZhongWilsonElimelech2020, RoyShenJoshiSongTuChowdhuryYeLiRenKumarAksimentievZeng2021, ShenFeiZhongFanSunHuGongCzajkowskyShao2021}.
Under optimized experimental conditions, carbon nanotube porins (CNTPs) embedded in lipid membranes allow the rapid permeation of water comparable to AQPs~\cite{TunuguntlaHenleyYaoPhamWanunuNoy2017, LiLiAydinQuanChenYaoZhanChenPhamNoy2020}.
A narrow radius in the subnanometer range of $\sim \! 0.8$~nm strictly rejects ions.
Another method involves cluster-forming peptides in lipid membranes which form an effective water pathway, resulting in fast and selective water permeation through water-wire networks~\cite{SongJoshiChowdhuryNajemShenLangHendersonTuFarellPitzMaranasCremerHickeySarlesHouAksimentievKumar2020, RoyShenJoshiSongTuChowdhuryYeLiRenKumarAksimentievZeng2021}.
For water purification, a tradeoff exists between permeability and selectivity~\cite{ParkKamcevRobesonElimelechFreeman2017}, i.e., a channel with a narrower radius rejects ion permeation but reduces water permeation.
Although the selectivity of artificial water channels has been actively investigated under osmotic gradients, unidirectional water transport without osmotic pressure or under inverse osmotic pressure has not been implemented.

Molecular dynamics (MD) simulations are useful for understanding and designing water behavior in confined nanoscale systems~\cite{HummerRasaiahNoworyta2001, SriramanKevrekidisHummer2005, Corry2011, Garc'ia-FandinoSansom2012, WinartoTakaiwaYamamotoYasuoka2015a, ZhouZhu2018, WinartoYamamotoYasuoka2019, KlesseTuckerSansom2020, AraiKoishiEbisuzaki2021}.
It has been shown that time-dependent deformation of CNTPs causes unidirectional water flow without osmotic pressure~\cite{LuNieWuZhouKouXuLiu2012, KouZhouLuXuWuFan2012, ZhouWuKouNieYangLu2013, ZhouZhu2018}, i.e., directional flow can be generated by introducing external energy as thermal fluctuations.
For technical implementations, the wettability or pore size within channels can be controlled by time-dependent external stimuli~\cite{LiuWangXieJuChu2016, ZhuWangTianJiang2019}, e.g., electric fields~\cite{RinneGekleBonthuisNetz2012, XiaoZhouKongXieLiZhangWenJiang2016, KlesseTuckerSansom2020}, light irradiation~\cite{CaiMaHaoSunGuoXuXuKuang2021}, electromagnetic fields~\cite{LiChangZhuSunFan2021}, and pH-dependent fields~\cite{ZhangHouZengYangLiYanTianJiang2013}.
However, how can water permeability be effectively controlled by external stimuli?

The permeable and impermeable states of hydrophobic ion channels depend on the wet and dewet states in a channel gate region, respectively~\cite{RaoKlesseStansfeldTuckerSansom2019, KlesseTuckerSansom2020}.
The transition between wetting and dewetting depends on the pore radius and works as an energy barrier to water and ion permeation~\cite{BecksteinBigginSansom2001, BecksteinSansom2003}.
Within a rigid nanotube, the transition generates a drift-like motion of water along the channel axis~\cite{HummerRasaiahNoworyta2001, SriramanKevrekidisHummer2005}.
AQPs also exhibit gating transition between permeable and impermeable states~\cite{KaptanAssentoftSchneiderFentonDeitmerMacAulayGroot2015, LindahlGourdonAnderssonHess2018}, where the pore radius exhibits fluctuations in the order of a few angstroms that follow $1/f$ noise attributed to memory-dependent dynamics~\cite{YamamotoAkimotoHiranoYasuiYasuoka2014}.
In other words, conformational fluctuation of biological channels is strongly related to the amount of transport.
Moreover, biological channels, e.g., voltage-gated ion channels, display hysteresis in the open/close transition of gating dynamics~\cite{MaennikkoePandeyLarssonElinder2005, TilegenovaCortesCuello2017}.
Hysteresis is a memory-dependent, non-linear behavior observed in various phenomena and incorporated in electronics, known as Schmitt trigger.
The current state is affected by the past states, which is important for resisting unwarranted noise.
The physiological role of hysteresis has been discussed as the conformational stability of permeable or nonpermeable states, leading to effective conduction~\cite{Villalba-Galea2017}.
Thus, we speculate that water conductivity could be effectively regulated by a memory-dependent fluctuation of the channel radius.
In other words, a memory-dependent noise-induced transition between the wet and dewet states could lead to a high transport ratio and capability to overcome the osmotic pressure.

In this study, we propose a design template of a ``water pump'' for extracting the work of directional water transport from thermal fluctuation.
We show that spatially asymmetric thermal fluctuations, i.e. $1/f^\alpha$ noise, imposed on the channel radius cause unidirectional water flow without osmotic pressure or a density gradient.
The proposed pump system achieves nanoscale water transport using a new principle based on the hysteresis, induced by wettability inside the channel.
The thermodynamics of the system can be explained by the inverse Carnot cycle with theoretical maximum efficiency, reminiscent of heat engines and pumps for energy conversion and transportation.

\section*{Results}
\subsection*{Unidirectional water transport through an active pump}
Figure~1A shows an active water pump system.
A membrane comprising two parallel sheets and a nanochannel was placed between two reservoirs with a periodic boundary condition, i.e., no osmotic pressure exists in the system.
The diameter of the channel was set to 1.25~nm to allow water molecules to pass in a single-file manner (Fig.~1B).
To simulate water transport by the active pump, the radius $R(t)$ of the left half of the nanochannel was fluctuated according to the following equation:
\begin{equation}
  \label{eq:random_noise}
           R(t) = R_0 + A \xi(t),
\end{equation}
where $t$ is time, $R_0$ is the average radius, and $A$ is the noise amplitude.
In the simulations, three types of random or correlated noises $\xi(t)$, i.e. white Gaussian noise, $1/f$ (pink) noise, and $1/f^2$ (Brownian) noise, were employed with the same mean and variance (Figs.~1C and 1D).
The $1/f^\alpha$ noises were generated by the following method~\cite{Goychuk2009, YamamotoAkimotoHiranoYasuiYasuoka2014}.
Here, $R_0$ and $A$ were set to 0.62 nm and 0.19, respectively.
These values were the most efficient values for water transportation in this model.
Furthermore, note that we confirmed that the water transport behavior does not significantly change with these parameters (Fig.~S2).

Initially, the pump system was equilibrated without any radius fluctuations (passive state, see supporting movie S1) for 1.0 $\mu$s, and then, the system was switched to an active (fluctuating) state for 1.8 $\mu$s to generate water flow by the active pump.
Spatially asymmetric thermal fluctuations imposed on the channel radius cause unidirectional water flows from the left side to the right side in the condition of no osmotic pressure in the reservoirs.
The cumulative number ($N_\mathrm{w}$) of water molecules passing from $x_\mathrm{left}$ to $x_\mathrm{right}$ was counted (Figs.~1B and 2A).
When the channel radius was fluctuated according to Brownian or pink noises, a stable water flow was generated by the nanochannel (Fig.~2A and SI movies S2 and S3). 
We estimated the transport rate $J$, which is defined as the number of water molecules passing through the channel over 1~ns, to estimate the water transport ability.
The pink noise, $J$, was approximately 1.0 ns$^{-1}$, which is the same as that of AQPs~\cite{ZeidelAmbudkarSmithAgre1992, BorgniaNielsenEngelAgre1999}.
Interestingly, $J$ of Brownian noise was approximately three times higher than that of pink noise.
In contrast, the channel radius fluctuation given by white noise generates little water flow (Supporting movie S4).

The dependency of the flow of water on different types of noises is related to the wetting and drying conditions inside the channel.
Figure~2B shows the number density profiles of water molecules inside the nanochannel during the passive and active states. 
In the passive state, the nanochannel is filled with water molecules, and several peaks exist along the $x$-direction.
This means that water molecules are trapped in the channel at equal intervals (Fig.~2C).
Switching to the active state dramatically changes the density profile.
At the white noise, the density of the left half ($x$ $<$ 0) becomes almost zero, i.e., the channel is dried at the radius fluctuating region (Fig.~2C).
Because of the high frequency component of radius variation, water molecules cannot penetrate the channel.
The densities with pink and Brownian noises are lower in most regions than those in the passive state; however, the density with Brownian noise was higher than that with pink noise overall.

\subsection*{Mechanism of unidirectional water transport}
The direction of flow is determined by the potential change in the fluctuating area.
When the nanochannel radius is sufficiently large, the water molecules are almost uniformly distributed in the channels (Figs.~2C and 2D), i.e., the potential energy is flat throughout the channel (Fig.~2E).
In the contraction process, the potential energy in the left half of the channel increases as the channel is narrowed.
Thus, the water molecules located near the middle of the channel ($x \sim 0$~nm) receive a rightward force because of the effect of the formation of a potential gradient.
Then, the water molecules in the right half of the channel ($x_\mathrm{center} < x < x_\mathrm{right}$) are pushed to the right reservoir along the gradient with a potential difference.
In contrast, it becomes difficult for water to enter the channel near the left terminal ($x \sim -3$~nm) because the height of the potential energy increases as the radius narrows.
The potential gradient to the right formed at the left terminal also leads to flow to the right.
Therefore, clearly, the potential gradient formed at the boundary between the fluctuating and the non-fluctuating regions leads to unidirectional flow through the channel.

\begin{figure}
\centering
\includegraphics[width=75 mm,bb= 0 0 703 414]{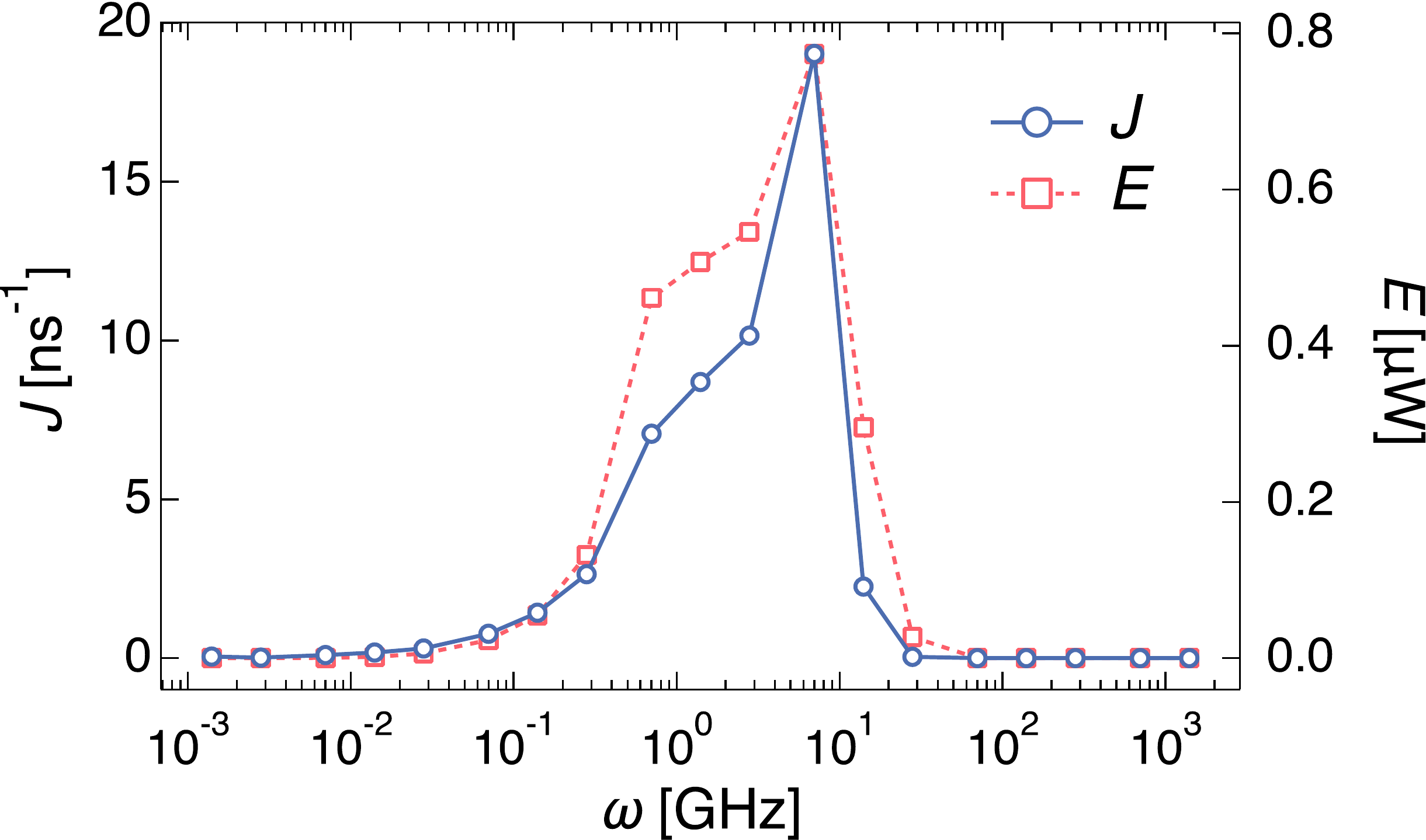}
\caption{Frequency dependence of water molecule transport as the nanochannel radius changes according to the sine wave ($R = A \sin(\omega t) + R_0$).
The left and right perpendicular axes represent the transport rate $J$ in ns$^{-1}$ and the work rate $E$ in $\rm \mu$W, respectively.}
\label{fig03}
\end{figure}

\subsection*{Frequency dependence of the water flux}
To investigate the dependency of the cumulative water flow on the different types of noises, we analyzed the water flow on each frequency.
An additional set of simulations was performed in which the channel radius $R(t)$ was changed with a certain frequency $\omega$, $R(t) = R_0 + A \sin \left( \omega t \right) $.
Different water transports with different $\omega$ values are shown in SI movie S5, where the values of $A$ and $R_0$ are the same as those in the previous simulations with different noises $\xi(t)$.
Figure~3 shows dependence of $\omega$ on the water transport rate $J$.
When $\omega < 10^{-1}$~GHz, almost no water transport occurs, i.e. $J \sim 0$.
In the larger frequency region, $J$ gradually increases and reaches a maximum value ($J \sim 19.02$~$\mathrm{ns}^{-1}$) at $\omega = 7.0$~GHz.
After that, $J$ rapidly decreases and becomes almost zero again at $\omega > 10^1$~GHz.
In other words, the passage of water molecules can be observed only in a certain range of frequencies ($10^{-1} < \omega < 10$~GHz).
To demonstrate the validity of the coarse-grained MD simulations, all-atom MD simulations of similar systems were performed, as shown in Fig.~S3A.
A similar dependence of the water transport rate on frequency with a single peak was observed in the all-atom MD simulations (Fig.~S3B).

\subsection*{Hysteresis in water transport by active pump}
Next, we monitored the changes in $N_\mathrm{L}$ during the oscillation of the channel radius.
Figure~4A shows the relationship between $N_\mathrm{L}$ and $R$ for some representative $\omega$ values.
The lower and upper arcs of the ellipse correspond to the expansion and contraction processes, respectively.
The black dashed curve in Fig.~4A represents $N_\mathrm{L}$ in the equilibrium state at $R$, which is called the ``static state.'' 
In the static state, for $R <$ 0.6 nm, no water can exist in the channel because the channel width is narrower than the size of the water molecules themselves.
When $R$ exceeds 0.58 nm, water begins to enter the channel, and $N_\mathrm{L}$ increases with $N_\mathrm{L}$ $\propto$ $R^2$ as $R$ increases further. 
This is because $N_\mathrm{L}$ is proportional to the volume inside the channel $V$ = $\pi R^2 L$ with a constant channel length $L$.

The $N_\mathrm{L}$--$R$ curves exhibit a rate-dependent hysteresis on $\omega$.
When the frequency is very low ($\omega = 7 \times 10^{-3}$ GHz), a state called the ``quasi-static state,'' the change in $N_\mathrm{L}$ agrees well with the static state. 
The water molecules begin to penetrate the channel at the radius of $R \sim 0.6$~nm in the expansion process.
After that, $N_\mathrm{L}$ rapidly increases with increasing $R$, and the nanochannel is filled with $\sim 10$ water molecules at the widest radius ($R \sim 0.7$~nm). 
Then, it switches to a contraction process in which $R$ gradually decreases.
The water molecules are completely excreted from the channel through almost the same path as the expansion process in the hysteresis curve.
At $\omega = 7 \times 10^{-1}$~GHz, the change in $N_\mathrm{L}$ significantly deviates from the static and quasi-static states. 
When $R$ is small, a few water molecules are not discharged from the channel and remain inside the channel.
Furthermore, a large hysteresis effect is observed when the expansion and contraction processes take completely different paths.
As $\omega$ increases to 7.0 GHz, $N_\mathrm{L}$ tends to decrease overall.
Thus, a flat shape is drawn in the $N_\mathrm{L}$--$R$ plane.
Furthermore, hysteresis also appears in all regions of $R$.
We call these two driving states the ``pump state''.
As $\omega$ increases further ($\omega = 7 \times 10^{1}$~GHz), no water molecules enter the channel; this state is called the ``closed state''.

\begin{figure}
\centering
\includegraphics[width=75 mm,bb= 0 0 642 837]{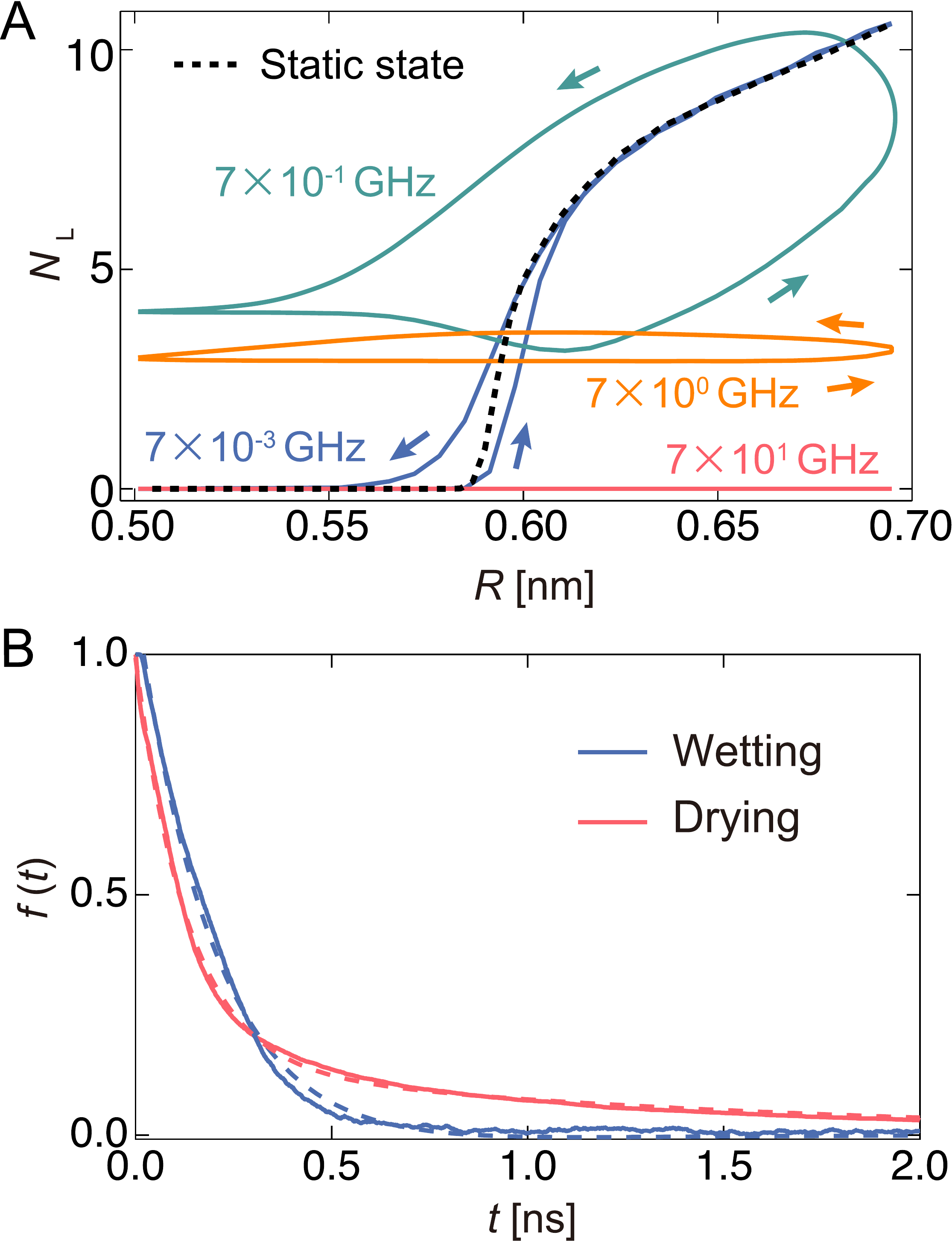}
\caption{Analysis of the wetting/drying dynamics in the active pump.
(A)~Hysteresis curves when the channel radius is controlled according to a sine wave.
The perpendicular and horizontal axes represent the number of water molecules, $N_\mathrm{L}$, located in the left half of the channel and the channel radius, $R$, respectively.
The black dashed curve denotes $N_\mathrm{L}$ in the equilibrium state at $R$.
(B)~The relaxation curves of water molecules inside the nanochannel after instantaneous expansion (wetting) or contraction (drying) of the channel radius.}
\label{fig04}
\end{figure}

\subsection*{Origin of hysteresis and relaxation times of the wetting/drying states inside the channel}
A clear hysteresis was observed during radial expansion and contraction in a certain frequency range (pump state). 
The hysteresis originates from the relationship between the period of expansion and contraction of the pump and the relaxation time of the wetting/drying states inside the channel.
To estimate the relaxation time using this model, we monitored the temporal change in the number of water molecules $N_\mathrm{L}$ in the left half of the channel when $R$ was instantly changed from the equilibrated wet ($R=0.69$~nm) or dry ($R=0.57$~nm) states to another state (Fig.~S4).
During the drying process, $N_\mathrm{L}$ rapidly decreased by 0.2 ns and then started slowly decreasing, reaching almost zero at approximately 10~ns.
This may correspond to two modes: one is extrusion by the nanochannel wall, and the other is escape by its own thermal diffusion. 
In the wetting process, water molecules rapidly fill the channel, and the steady state is reached in approximately 1.0~ns.
The number of water molecules in the steady state is defined as $N_\mathrm{C}$.
To calculate the relaxation times of the wetting and drying processes, the time variation in $f(t)$ is shown in Figure~4B. 
Here, $f(t)$ is given by $N_\mathrm{L}(t)/N_\mathrm{L}(t_0)$ in the drying process and $[N_\mathrm{C} - N_\mathrm{L}(t)]/[N_\mathrm{C} - N_\mathrm{L}(t_0)]$ in the wetting process. 
The wetting and drying data are fitted to a single exponential curve, $f(t) = A' \exp \left( -t / \tau_\mathrm{wet} \right)$, and double exponential curve, $ f(t) = A'' \exp \left( -t/\tau_\mathrm{dry1} \right) + (1-A'') \exp \left( -t /\tau_\mathrm{dry2} \right)$, respectively.
Thus, $\tau_\mathrm{wet}$, $\tau_\mathrm{dry1}$, and $\tau_\mathrm{dry2}$ are 0.348, 0.123, and 2.482~ns respectively.
Here, $\tau_\mathrm{dry1}$ and $\tau_\mathrm{dry2}$ correspond to the relaxation time forced out by tube contraction and the relaxation time spontaneously released by thermal diffusion, respectively.
In this pump system, the water molecules are expelled at a faster rate than the rate with which they fill the channel ($\tau_\mathrm{dry1} < \tau_\mathrm{wet}$).
In the drying process, water molecules are pushed out by the pressure ($p_\mathrm{wall}$) of the wall due to radial contraction, whereas in the wetting process, the reservoir is filled by the pressure ($p_0$) of the bulk water.

\begin{table*}
\centering
  \caption{Data list of half period $P_\mathrm{half} = \pi/\omega$, ratio of the relaxation time $\tau$ to $P_\mathrm{half}$, and transport rate $J$ for typical frequencies $\omega$.}
  \label{hyst_curve}
  \begin{tabular}{lrrrrrc}
     $\omega$ [GHz] & $P_\mathrm{half}$ [ns] & $P_\mathrm{half}/\tau_\mathrm{dry1}$ & $P_\mathrm{half}/\tau_\mathrm{wet}$ & $P_\mathrm{half}/\tau_\mathrm{dry2}$ & $J$ [ns$^{-1}$] & State\\
	 \hline
    7$\times 10^{-3}$	&	449		&	3636				&	1288				&	181					&	0.09	&	Quasi-static	\\
    7$\times 10^{-1}$	&	4.49	&	36.4				&	12.9				&	1.81				&	7.06	&	Pump			\\
    7$\times 10^{0}$	&	0.449	&	3.64				&	1.29				&	0.181	&	19.02	&	Pump			\\
    7$\times 10^{1}$	&	0.0449	&	0.364	&	0.129	&	0.0181	&	0.00	&	Closed			\\
	\hline
  \end{tabular}
\end{table*}

The upper and lower limits of the frequency for the hysteresis are determined by the relaxation times, $\tau_\mathrm{wet}$, $\tau_\mathrm{dry1}$, and $\tau_\mathrm{dry2}$.
Table~1 shows half periods ($P_\mathrm{half} = \pi/\omega$), ratios of the relaxation time to $P_\mathrm{half}$, transport rates $J$, and the corresponding pump states for the typical frequencies.
We focus on the half-period here because half of a cycle is a contraction process and the other half is an expansion process.
The slowest relaxation time among the three relaxation times is $\tau_\mathrm{dry2}$, which is the characteristic time required for water molecules to be ejected from the channel by thermal diffusion.
In the quasi-static state, the period of the radial change is approximately 180 times slower than $\tau_\mathrm{dry2}$.
In other words, water molecules are spontaneously discharged into the reservoir by diffusion without being pushed out by the channel wall. 
Therefore, almost no directionality exists in the outflow of water molecules, and $J$ is extremely low ($\sim$ 0.09 ns$^{-1}$).
The fastest relaxation time, in contrast, is $\tau_\mathrm{dry1}$, which is the characteristic time of being pushed out of the channel by the wall surface.
In the closed state, the period of radial change $P_\mathrm{half}$ is less than half of $\tau_\mathrm{dry1}$; in addition, $P_\mathrm{half}/\tau_\mathrm{wet}$ is only approximately 0.129. 
Before water can enter the channel through expansion, the channel closes by switching to the contraction process.
Thus, once water is drained out, it cannot enter the channel again, and $J$ becomes zero.
In the pump state, where hysteresis is evident, the period of the change in $R$ is in the appropriate balance between $\tau_\mathrm{wet}$ and $\tau_\mathrm{dry1}$, and $\tau_\mathrm{dry2}$.
In other words, water molecules supplied from the reservoir during the expansion process can be reliably pushed in one direction during the contraction process. 
Moreover, the rate of non-directional water movement due to thermal diffusion is small. 
The value of $J$ at $\omega = 7$ GHz is larger than that at $\omega = 70$ GHz because of $P_\mathrm{half}/\tau_\mathrm{dry2}$.
Thus, continuous work is achieved in the pump state.

\begin{figure}
\centering
\includegraphics[width=75 mm,bb= 0 0 607 633]{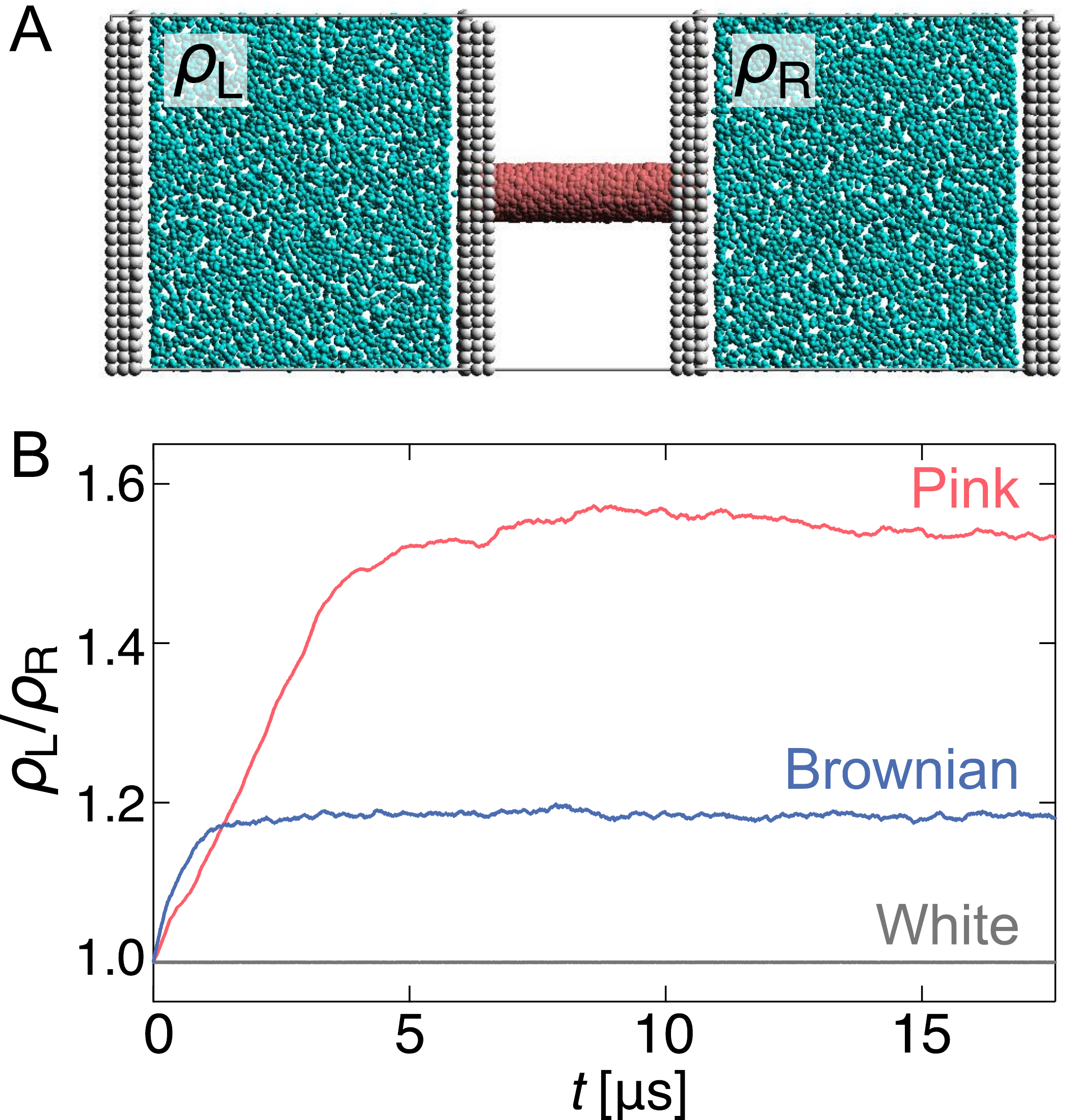}
\caption{Capability to overcome osmotic pressure.
(A)~Snapshots of a separated active pump system.
(B)~Time series of the density difference ($\rho_\mathrm{L}/\rho_\mathrm{R}$) produced by active pumps with different noises.
Here, $\rho_\mathrm{L}$ and $\rho_\mathrm{R}$ represent the number densities of the left and right reservoirs, respectively.
Each curve is plotted based on the averaged data over 10 independent simulations.}
\label{fig05}
\end{figure}

\subsection*{Capability to overcome osmotic pressure}
Finally, we performed simulations to investigate whether this pump system overcomes the osmotic pressure and can transport water. 
First, a separated pump system was prepared in which the reservoirs on either side of the nanochannel were unconnected and equilibrated (Fig.~5A).
Then, after inducing $R$ fluctuating to the channel, the density change of each reservoir ($\rho_\mathrm{L}/\rho_\mathrm{R}$) was monitored (Fig.~5B).

As expected, when $R$ fluctuation follows white Gaussian noise, $\rho_\mathrm{L}/\rho_\mathrm{R}$ remains 1.0 because the water molecules do not pass through the nanochannel. 
In contrast, for pink and Brownian noises, $\rho_\mathrm{L}/\rho_\mathrm{R}$ increases with time against the osmotic pressure and reaches a steady state with a certain density difference. 
The transport rate of the Brownian noise pump is higher than that of the pink noise pump; however, it is balanced at a steady state, $\rho_\mathrm{L}/\rho_\mathrm{R} \sim 1.18$, after $t = 2$~$\mu$s.
The pink noise pump has a higher capability to overcome the osmotic pressure, and a density difference of approximately 1.53 times is achieved in about 10~$\mu$s.

Why is it possible to achieve a higher density difference in pink noise, even though the transport rate is slower than that in Brownian noise? 
This can be explained by the frequency components in the pink noise.
As shown in the previous section, this pump system has a characteristic relaxation time.
Water transport efficiently occurs when the frequency range of the channel radius is from $7 \times 10^{-1}$ to $10^1$~GHz as shown in Fig.~3.
As shown in Fig.~1D, the spectral density of pink noise is higher than that of Brownian noise under most frequencies. 
Hence, water molecules entering from the reservoir are returned to the original reservoir by the slow radial change in Brownian noise, whereas pink noise can efficiently send them to the channel side.
The thermal fluctuation of water channels such as AQPs follows $1/f$ noise~\cite{YamamotoAkimotoHiranoYasuiYasuoka2014}. 
Our results help explain the inevitability of $1/f$ fluctuations in biological water channels.

\subsection*{Discussion}
In summary, we have provided in silico evidence for an active water transport pump driven by thermal fluctuations on pore radius. 
Our proposed pump is independent of the external environment and can transport water while overcoming the density gradients and osmotic pressure.
The proposed pump is a nanoscale energy conversion system that corresponds to a reversed Carnot cycle, which utilizes hysteresis in the expansion and contraction processes to transport water.

A hysteresis property has already been used in practical applications as an energy conversion device.
For example, a Schmitt-trigger oscillator is a device that converts electrical energy into vibrational energy.
The Schmitt trigger is an electronic circuit characterized by hysteresis in the output state in response to changes in the input voltage. 
By setting two thresholds for the input voltage, the output voltage can be switched with the delay time with a change in the input voltage.
In our proposed pump system, the input and output voltages correspond to the change in the radius of the nanochannel and the number of water molecules in the channel, respectively.
Because the filling and ejection of water molecules in the channel are not synchronous with the change in the channel radius, hysteresis (delay time) exists between the expansion and contraction processes.

Additionally, the hysteresis curves (Fig.~S5) represent thermodynamic cycles, indicating that the water pump can be regarded as a nanoscale heat pump. 
The processes of the water pump with constant frequency roughly correspond to the following four processes of the reversed Carnot cycle; (a)~adiabatic compression: contract without outflow, (b)~isothermal compression: contract with outflow, (c)~adiabatic expansion: expand without inflow, and (d)~isothermal expansion: expand with inflow, as shown in Fig.~S1 (Detailed description is written in SI).
The temperature in the water pump increases or decreases because of adiabatic processes caused by the radius change of the channel, which indices ejection or penetration of water molecules, respectively.
The flow of water molecules induced by this process corresponds to the heat flow in a heat pump, and the water pump can behave as the nanoscale heat pump that transforms the work of radius change into molecule flow. 

Using this hysteresis curve, we estimated the work rate ($E$) for each frequency. 
To estimate these as work dimensions, we converted $N_\mathrm{L}$ and $R$ to the pressure ($p$) and volume ($V$), respectively (Fig.~S5). 
In this study, $p$ was defined as $N_\mathrm{L}$ multiplied by the force in the $x$-direction that a water molecule in the channel receives from the wall when a stable water flow is generated, and then, divided by the cross-sectional area of the channel.
The volume of the left half of the channel was defined as $V$.
$E$ was estimated based on the area inside the $p$--$V$ hysteresis curve integrated and multiplied by $\omega$.
The right axis of Fig.~3 shows the frequency dependence of $E$.
A clear correlation between $J$ and $E$ was observed.
As mentioned earlier, the channel radius expands and contracts in the characteristic time between $\tau_\mathrm{wet}$ and $\tau_\mathrm{dry}$ in this frequency region, and water flow is generated by forcing the water out through the contraction process. 
This workload was shown to agree well with the water transport rate.

Additionally, this energy conversion system through nanotubes, similar to a reversed Carnot cycle, is driven by a radius change of only $0.2$~nm, as shown in Fig.~4A.
The size of this change was sufficiently large to be achieved by thermal fluctuations, and actual water channels such as AQPs can control the amount of water that could permeate according to this mechanism.

Moreover, we revealed that the structure fluctuations are a key parameter that affects water transport through the nanochannel. 
For water transport, the channel radius fluctuation should not follow white noise but should rather follow $1/f^\alpha$ noise. 
This is because certain frequencies that cause drying and wetting inside the channel are mostly contained in the $1/f^\alpha$ fluctuations.
This may explain why many biological channels exhibit memory-dependent dynamics, e.g., $1/f^\alpha$ noise and hysteresis, in structural fluctuations.
In terms of the technological implementation, the wettability or pore size of channels can be controlled by time-dependent external stimuli~\cite{LiuWangXieJuChu2016, ZhuWangTianJiang2019}.
In combination with these studies, we believe that the water pump system proposed here may be an effective device for low-energy artificial purification systems and for dealing with possible water shortages in the future.

\section*{Materials and Methods}
\subsection*{Coarse-grained molecular simulation}
In this study, dissipative particle dynamics (DPD) simulation~\cite{HoogerbruggeKoelman1992, GrootWarren1997} was adopted to reproduce the water transportation of active pumps at the molecular level. 
The DPD method is a coarse-grained molecular simulation method, which has proven to be a powerful tool for investigating fluid events occurring on a wide range of spatio-temporal scales compared to all-atom simulations such as a molecular dynamics (MD) method.
The DPD method is based on Newton's equation of motion, which is expressed as follows:
 \begin{equation}
	m_i \frac{d \textbf{v}_i}{dt} = {\textbf{f}_i} = \sum_{j \neq i} {\textbf{F}}_{ij}^\mathrm{C} + \sum_{j \neq i} {\textbf{F}}_{ij}^\mathrm{D} + \sum_{j \neq i} {\textbf{F}}_{ij}^\mathrm{R}\;\;, 
  \label{eq:eq_motion}
\end{equation}
where $i$ and $j$ are the particle index, $m$ is the particle mass, $\textbf{v}$ is velocity, and $\textbf{f}$ is force. 
Each DPD particle is subject to three types of forces (conservative ($\textbf{F}^{\rm{C}}$), dissipative ($\textbf{F}^{\rm{R}}$), and random ($\textbf{F}^{\rm{D}}$). 
 The conservative force is softly repulsive and is given by
 \begin{equation}
	{\textbf{F}}_{ij}^\mathrm{C} =
		\begin{cases}
			-a_{ij} \left( 1-\dfrac{ \left| \textbf{r}_{ij}\right|}{r_{\mathrm c}} \right) \textbf{n}_{ij}, & \left| \textbf{r}_{ij} \right| \leq r_{\mathrm c} \\ 
			                      \;\;\;\;\;\;\;\;\;\;\;\;\;\;\;0,	& \left| \textbf{r}_{ij} \right| > r_{\mathrm c}\;\;,
		\end{cases}
	\label{eq:FC}
\end{equation}
where $\textbf{r}$ is the position vector, $\textbf{r}_{ij} = \textbf{r}_{j} - \textbf{r}_{i}$, and $\textbf{n}_{ij} = \textbf{r}_{ij} / \left| \textbf{r}_{ij} \right|$. Here, $a_{ij}$ is a parameter that can determine the magnitude of the repulsive force between particles, and $r_{\mathrm c}$ is the cutoff distance. 
The dissipative and random forces are given by
\begin{equation}
	\label{eq:FD}
	{\textbf{F}}_{ij}^\mathrm{D} =
		\begin{cases}
			- \gamma \omega^{\mathrm D} \left( \left| \textbf{r}_{ij} \right| \right) \left( \textbf{n}_{ij} \cdot \textbf{v}_{ij} \right) \textbf{n}_{ij}, & \left| \textbf{r}_{ij} \right| \leq r_{\mathrm c}  \\ 
			                      \;\;\;\;\;\;\;\;\;\;\;\;\;\;\;0,	& \left| \textbf{r}_{ij} \right| > r_{\mathrm c}  
		\end{cases}
\end{equation}
and
\begin{equation}
	\label{eq:FR}
	{\textbf{F}}_{ij}^\mathrm{R} =
		\begin{cases}
			\sigma \omega^{\mathrm R} \left( \left| \textbf{r}_{ij}\right| \right) \zeta_{ij} \Delta t^{-1/2} \textbf{n}_{ij}, & \left| \textbf{r}_{ij} \right| \leq r_{\mathrm c}\\ 
			                      \;\;\;\;\;\;\;\;\;\;\;\;\;\;\;0,	& \left| \textbf{r}_{ij} \right| > r_{\mathrm c} 
		\end{cases}
\end{equation}
respectively, 
where $\textbf{v}_{ij} = \textbf{v}_{j} - \textbf{v}_{i}$, $\sigma$ is the noise parameter, $\gamma$ is the friction parameter, and $\zeta_{ij}$ is the random number based on the Gaussian distribution.
Here, $\omega^{\mathrm R}$ and $\omega^{\mathrm D}$ are $r$--dependent weight functions and are given as follows:
 \begin{equation}
  \label{eq:w_func}
	\omega^\mathrm{D} \left( r \right) = \left[ \omega^\mathrm{R} \left( r \right) \right]^2 = 
		\begin{cases}
			\left[ 1 - \dfrac{\left| \textbf{r}_{ij} \right|}{r_c}\right]^2,	& \left| \textbf{r}_{ij} \right| \leq r_{\mathrm c} \\ 
			                      \;\;\;\;0,	& \left| \textbf{r}_{ij} \right| > r_{\mathrm c} \;\; .
		\end{cases}
\end{equation}
The temperature is controlled by a combination of dissipative and random forces.
The noise parameter $\sigma$ and friction parameter $\gamma$ are associated by the fluctuation-dissipation theorem given by the following equation:
 \begin{equation}
  \label{eq:fd_theory}
	\sigma ^2 = 2 \gamma k_\mathrm{B} T,
\end{equation}
where $k_{\mathrm B}$ is the Boltzmann constant and $T$ is the temperature.

The results of DPD simulations are commonly reported in reduced units.
The DPD unit of length is the cutoff distance ($r_\mathrm{c}$), the unit of mass is the particle mass ($m$), and the unit of energy is ($k_{\mathrm B}T$). In this study, we adopted the method developed by Groot and Rabone\cite{GrootRabone2001} for scaling the length and time units since the system is dominated by water behavior.

Based on our previous study~\cite{AraiKoishiEbisuzaki2021}, the active water pump system comprises a nanochannel (6.46 nm $<$ $x$ $<$ 12.92 nm), a water reservoir ($x$ $<$ 6.46 nm or $x$ $>$ 12.92 nm), and partition walls ($x$ $\sim$ 6.46 and 12.92 nm) between the nanochannel and the reservoir.
The length and diameter of the nanochannel were set to 6.46 nm and 1.25 nm, respectively.
A water molecule is represented by a single DPD particle as is often used in this technique.
Table~2 shows the interactions between any two particles.
These interaction parameters are set based on our previous motor protein systems~\cite{AraiYasuokaKoishiEbisuzaki2010, AraiYasuokaKoishiEbisuzakiZeng2013}.
The total number of particles in this model is 14,105, and the number of water and nanochannel beads is 9,083 and 1,476, respectively; the remaining particles are partitions.
The volume of the simulation box is $19.39 \times 6.46 \times 9.69$~nm$^3$.

By connecting the nanochannel and wall particles to their initial positions using a harmonic spring, the thermal fluctuation of the solid can be reproduced. The spring constant is set to 10$^4$ $k_\mathrm{B}T/r_\mathrm{c}^2$~\cite{AraiKoishiEbisuzaki2021}.
The noise parameter $\sigma$ is set as 3.0, and the friction parameter $\gamma$ as 4.5.
The periodic boundary condition is applied in all three dimensions.
All simulations are performed in the constant-volume and constant-temperature ensemble.

\subsection*{All-atom MD simulations}
We also performed MD simulations to study the behaviors of water molecules in the nanoscale active pump whose radius of the left half is changed with a certain frequency, \(R(t)=R_0+A\sin(\omega t)\).
The settings of the MD simulation were almost the same as those in our previous study~\cite{AraiKoishiEbisuzaki2021} in which the system contains water molecules, two graphite plates, and a carbon nano tube (CNT) in a periodic boundary simulation box with size of $8.64 \times 5.41 \times 5.12$~nm$^3$ as shown in Fig.~S3A. 
The two parallel graphite plates were placed to separate the nanochannel region from the water reservoir region as well as the DPD system. 
The chirality and diameter of the CNT are $(6, 6)$ and 0.81~nm, respectively~\cite{ObaOkadaMaruyama2005}.
In the nanochannel region, the CNT is located between the graphite plates. Each plate has a hole at the place where CNT is in contact so that the water molecules in the reservoir can flow into the CNT. The direction of CNT is parallel to the \(x\)-axis, and the length of the CNT is 4.0~nm.
The \(x\)-direction size of the reservoir is 4.64~nm, and the \(y\) and \(z\) sizes are the same as those of the simulation box.
The reservoir is separately located on the left and right sides of the simulation box; however, they are connected through a periodic boundary condition. 
The LJ parameters $\sigma=0.34$~nm and $\varepsilon=0.2328$~kJ/mol are used for graphite carbon atoms~\cite{BhethanabotlaSteele1987}.
During the simulation, the carbon atoms are fixed.

A rigid-body model of water, the SPC/E model~\cite{BerendsenGrigeraStraatsma1987}, is employed for water molecules.
The potential function of the SPC/E model includes two terms---a Coulomb term and a Lennard-Jones (LJ) term. The long-range charge--charge interactions between the water molecules can be calculated using the particle-mesh Ewald (PME) method\cite{EssmannPereraBerkowitzDardenLeePedersen1995}. The cut-off length of the Ewald real part and LJ interactions is 1.2~nm.
Rotational motion of water molecules is calculated using the symplectic quaternion integrator~\cite{MillerIIIEleftheriouPattnaikNdirangoNewnsMartyna2002}.
The MD time step is set at 2~fs, and the total sampling simulation time is 10~ns. 
An MD simulation program previously developed by our group~\cite{KoishiTakeichi2015} was employed to perform all simulations.

Water molecules were set using a simple cubic lattice in the reservoir region. 
The total number of water molecules was 4284. 
The system was equilibrated for another 1.0~ns at $T=298$~K. 
These equilibration runs were performed in a constant-volume and constant-energy ($NVE$) ensemble. 
The sampling runs for 10~ns to estimate water flow were performed in a constant-volume and constant-temperature ($NVT$) ensemble with the Nos\'e-Poinca\'e method \cite{Nose2001}. 
The averaged water density in the reservoir was 1.06~g/cm$^3$ in the sampling runs.

\begin{table}
\centering
  \caption{Interaction parameters of pair particles $a_{ij}$ ($k_\mathrm{B}T/r_\mathrm{c}$ unit).}
  \label{parameter_bead}
  \begin{tabular}{lrrr}
     & Channel & Partition & Water \\
	 \hline
    Channel		&	0	&	0	&	100	\\
    Partition	&	0	&	0	&	50	\\
    Water		&	100	&	50	&	25	\\
	\hline
  \end{tabular}
\end{table}

\subsection*{Acknowledgements}
This work was supported in part by the Very Generous Foundation.
Author Contributions: T.E. conceptualized the research.
N.A. and T.K. conceived the research.
N.A., E.Y. and K.Y. designed the analysis.
N.A. and T.K. conducted the analyses.
All authors discussed the results.
N.A. and E.Y. mainly wrote the manuscript.


\begin{thebibliography}{66}%
\makeatletter
\providecommand \@ifxundefined [1]{%
 \@ifx{#1\undefined}
}%
\providecommand \@ifnum [1]{%
 \ifnum #1\expandafter \@firstoftwo
 \else \expandafter \@secondoftwo
 \fi
}%
\providecommand \@ifx [1]{%
 \ifx #1\expandafter \@firstoftwo
 \else \expandafter \@secondoftwo
 \fi
}%
\providecommand \natexlab [1]{#1}%
%
\providecommand \bibnamefont  [1]{#1}%
\providecommand \bibfnamefont [1]{#1}%
\providecommand \citenamefont [1]{#1}%
\providecommand \href@noop [0]{\@secondoftwo}%
\providecommand \href [0]{\begingroup \@sanitize@url \@href}%
\providecommand \@href[1]{\@@startlink{#1}\@@href}%
\providecommand \@@href[1]{\endgroup#1\@@endlink}%
\providecommand \@sanitize@url [0]{\catcode `\\12\catcode `\$12\catcode
  `\&12\catcode `\#12\catcode `\^12\catcode `\_12\catcode `\%12\relax}%
\providecommand \@@startlink[1]{}%
\providecommand \@@endlink[0]{}%
\providecommand \url  [0]{\begingroup\@sanitize@url \@url }%
\providecommand \@url [1]{\endgroup\@href {#1}{\urlprefix }}%
\providecommand \urlprefix  [0]{URL }%
%
\providecommand \doibase [0]{http://dx.doi.org/}%
\providecommand \selectlanguage [0]{\@gobble}%
\providecommand \bibinfo  [0]{\@secondoftwo}%
\providecommand \bibfield  [0]{\@secondoftwo}%
%
\providecommand \BibitemOpen [0]{}%
%
%
%
\providecommand \BibitemShut  [1]{\csname bibitem#1\endcsname}%
\let\auto@bib@innerbib\@empty
\bibitem [{\citenamefont {Werber}\ \emph {et~al.}(2016)\citenamefont {Werber},
  \citenamefont {Osuji},\ and\ \citenamefont
  {Elimelech}}]{WerberOsujiElimelech2016}%
  \BibitemOpen
  \bibfield  {author} {\bibinfo {author} {\bibfnamefont {J.~R.}\ \bibnamefont
  {Werber}}, \bibinfo {author} {\bibfnamefont {C.~O.}\ \bibnamefont {Osuji}}, \
  and\ \bibinfo {author} {\bibfnamefont {M.}~\bibnamefont {Elimelech}},\
  }\href@noop {} {\bibfield  {journal} {\bibinfo  {journal} {Nat. Rev. Mater.}\
  }\textbf {\bibinfo {volume} {1}},\ \bibinfo {pages} {16018} (\bibinfo {year}
  {2016})}\BibitemShut {NoStop}%
\bibitem [{\citenamefont {Shen}\ \emph
  {et~al.}(2021{\natexlab{a}})\citenamefont {Shen}, \citenamefont {Liu},
  \citenamefont {Han},\ and\ \citenamefont {Jin}}]{ShenLiuHanJin2021}%
  \BibitemOpen
  \bibfield  {author} {\bibinfo {author} {\bibfnamefont {J.}~\bibnamefont
  {Shen}}, \bibinfo {author} {\bibfnamefont {G.}~\bibnamefont {Liu}}, \bibinfo
  {author} {\bibfnamefont {Y.}~\bibnamefont {Han}}, \ and\ \bibinfo {author}
  {\bibfnamefont {W.}~\bibnamefont {Jin}},\ }\href@noop {} {\bibfield
  {journal} {\bibinfo  {journal} {Nat. Rev. Mater.}\ }\textbf {\bibinfo
  {volume} {6}},\ \bibinfo {pages} {294} (\bibinfo {year}
  {2021}{\natexlab{a}})}\BibitemShut {NoStop}%
\bibitem [{\citenamefont {Barboiu}\ and\ \citenamefont
  {Gilles}(2013)}]{BarboiuGilles2013}%
  \BibitemOpen
  \bibfield  {author} {\bibinfo {author} {\bibfnamefont {M.}~\bibnamefont
  {Barboiu}}\ and\ \bibinfo {author} {\bibfnamefont {A.}~\bibnamefont
  {Gilles}},\ }\href@noop {} {\bibfield  {journal} {\bibinfo  {journal} {Acc.
  Chem. Res.}\ }\textbf {\bibinfo {volume} {46}},\ \bibinfo {pages} {2814}
  (\bibinfo {year} {2013})}\BibitemShut {NoStop}%
\bibitem [{\citenamefont {Huo}\ and\ \citenamefont {Zeng}(2016)}]{HuoZeng2016}%
  \BibitemOpen
  \bibfield  {author} {\bibinfo {author} {\bibfnamefont {Y.}~\bibnamefont
  {Huo}}\ and\ \bibinfo {author} {\bibfnamefont {H.}~\bibnamefont {Zeng}},\
  }\href@noop {} {\bibfield  {journal} {\bibinfo  {journal} {Acc. Chem. Res.}\
  }\textbf {\bibinfo {volume} {49}},\ \bibinfo {pages} {922} (\bibinfo {year}
  {2016})}\BibitemShut {NoStop}%
\bibitem [{\citenamefont {Lynch}\ \emph {et~al.}(2020)\citenamefont {Lynch},
  \citenamefont {Rao},\ and\ \citenamefont {Sansom}}]{LynchRaoSansom2020}%
  \BibitemOpen
  \bibfield  {author} {\bibinfo {author} {\bibfnamefont {C.~I.}\ \bibnamefont
  {Lynch}}, \bibinfo {author} {\bibfnamefont {S.}~\bibnamefont {Rao}}, \ and\
  \bibinfo {author} {\bibfnamefont {M.~S.~P.}\ \bibnamefont {Sansom}},\
  }\href@noop {} {\bibfield  {journal} {\bibinfo  {journal} {Chem. Rev.}\
  }\textbf {\bibinfo {volume} {120}},\ \bibinfo {pages} {10298} (\bibinfo
  {year} {2020})}\BibitemShut {NoStop}%
\bibitem [{\citenamefont {Murata}\ \emph {et~al.}(2000)\citenamefont {Murata},
  \citenamefont {Mitsuoka}, \citenamefont {Hirai}, \citenamefont {Walz},
  \citenamefont {Agre}, \citenamefont {Heymann}, \citenamefont {Engel},\ and\
  \citenamefont
  {Fujiyoshi}}]{MurataMitsuokaHiraiWalzAgreHeymannEngelFujiyoshi2000}%
  \BibitemOpen
  \bibfield  {author} {\bibinfo {author} {\bibfnamefont {K.}~\bibnamefont
  {Murata}}, \bibinfo {author} {\bibfnamefont {K.}~\bibnamefont {Mitsuoka}},
  \bibinfo {author} {\bibfnamefont {T.}~\bibnamefont {Hirai}}, \bibinfo
  {author} {\bibfnamefont {T.}~\bibnamefont {Walz}}, \bibinfo {author}
  {\bibfnamefont {P.}~\bibnamefont {Agre}}, \bibinfo {author} {\bibfnamefont
  {J.~B.}\ \bibnamefont {Heymann}}, \bibinfo {author} {\bibfnamefont
  {A.}~\bibnamefont {Engel}}, \ and\ \bibinfo {author} {\bibfnamefont
  {Y.}~\bibnamefont {Fujiyoshi}},\ }\href@noop {} {\bibfield  {journal}
  {\bibinfo  {journal} {Nature}\ }\textbf {\bibinfo {volume} {407}},\ \bibinfo
  {pages} {599} (\bibinfo {year} {2000})}\BibitemShut {NoStop}%
\bibitem [{\citenamefont {Agre}\ \emph {et~al.}(2002)\citenamefont {Agre},
  \citenamefont {King}, \citenamefont {Yasui}, \citenamefont {Guggino},
  \citenamefont {Ottersen}, \citenamefont {Fujiyoshi}, \citenamefont {Engel},\
  and\ \citenamefont
  {Nielsen}}]{AgreKingYasuiGugginoOttersenFujiyoshiEngelNielsen2002}%
  \BibitemOpen
  \bibfield  {author} {\bibinfo {author} {\bibfnamefont {P.}~\bibnamefont
  {Agre}}, \bibinfo {author} {\bibfnamefont {L.~S.}\ \bibnamefont {King}},
  \bibinfo {author} {\bibfnamefont {M.}~\bibnamefont {Yasui}}, \bibinfo
  {author} {\bibfnamefont {W.~B.}\ \bibnamefont {Guggino}}, \bibinfo {author}
  {\bibfnamefont {O.~P.}\ \bibnamefont {Ottersen}}, \bibinfo {author}
  {\bibfnamefont {Y.}~\bibnamefont {Fujiyoshi}}, \bibinfo {author}
  {\bibfnamefont {A.}~\bibnamefont {Engel}}, \ and\ \bibinfo {author}
  {\bibfnamefont {S.}~\bibnamefont {Nielsen}},\ }\href@noop {} {\bibfield
  {journal} {\bibinfo  {journal} {J. Physiol.}\ }\textbf {\bibinfo {volume}
  {542}},\ \bibinfo {pages} {3} (\bibinfo {year} {2002})}\BibitemShut {NoStop}%
\bibitem [{\citenamefont {de~Groot}\ and\ \citenamefont
  {Grubm{\"u}ller}(2001)}]{GrootGrubmuller2001}%
  \BibitemOpen
  \bibfield  {author} {\bibinfo {author} {\bibfnamefont {B.~L.}\ \bibnamefont
  {de~Groot}}\ and\ \bibinfo {author} {\bibfnamefont {H.}~\bibnamefont
  {Grubm{\"u}ller}},\ }\href@noop {} {\bibfield  {journal} {\bibinfo  {journal}
  {Science}\ }\textbf {\bibinfo {volume} {294}},\ \bibinfo {pages} {2353}
  (\bibinfo {year} {2001})}\BibitemShut {NoStop}%
\bibitem [{\citenamefont {Savage}\ \emph {et~al.}(2010)\citenamefont {Savage},
  \citenamefont {O'Connell}, \citenamefont {Miercke}, \citenamefont
  {Finer-Moore},\ and\ \citenamefont
  {Stroud}}]{SavageOConnellMierckeFiner-MooreStroud2010}%
  \BibitemOpen
  \bibfield  {author} {\bibinfo {author} {\bibfnamefont {D.~F.}\ \bibnamefont
  {Savage}}, \bibinfo {author} {\bibfnamefont {J.~D.}\ \bibnamefont
  {O'Connell}}, \bibinfo {author} {\bibfnamefont {L.~J.~W.}\ \bibnamefont
  {Miercke}}, \bibinfo {author} {\bibfnamefont {J.}~\bibnamefont
  {Finer-Moore}}, \ and\ \bibinfo {author} {\bibfnamefont {R.~M.}\ \bibnamefont
  {Stroud}},\ }\href@noop {} {\bibfield  {journal} {\bibinfo  {journal} {Proc.
  Natl. Acad. Sci. USA}\ }\textbf {\bibinfo {volume} {107}},\ \bibinfo {pages}
  {17164} (\bibinfo {year} {2010})}\BibitemShut {NoStop}%
\bibitem [{\citenamefont {Gravelle}\ \emph {et~al.}(2013)\citenamefont
  {Gravelle}, \citenamefont {Joly}, \citenamefont {Detcheverry}, \citenamefont
  {Ybert}, \citenamefont {Cottin-Bizonne},\ and\ \citenamefont
  {Bocquet}}]{GravelleJolyDetcheverryYbertCottin-BizonneBocquet2013}%
  \BibitemOpen
  \bibfield  {author} {\bibinfo {author} {\bibfnamefont {S.}~\bibnamefont
  {Gravelle}}, \bibinfo {author} {\bibfnamefont {L.}~\bibnamefont {Joly}},
  \bibinfo {author} {\bibfnamefont {F.}~\bibnamefont {Detcheverry}}, \bibinfo
  {author} {\bibfnamefont {C.}~\bibnamefont {Ybert}}, \bibinfo {author}
  {\bibfnamefont {C.}~\bibnamefont {Cottin-Bizonne}}, \ and\ \bibinfo {author}
  {\bibfnamefont {L.}~\bibnamefont {Bocquet}},\ }\href@noop {} {\bibfield
  {journal} {\bibinfo  {journal} {Proc. Natl. Acad. Sci. USA}\ }\textbf
  {\bibinfo {volume} {110}},\ \bibinfo {pages} {16367} (\bibinfo {year}
  {2013})}\BibitemShut {NoStop}%
\bibitem [{\citenamefont {Eriksson}\ \emph {et~al.}(2013)\citenamefont
  {Eriksson}, \citenamefont {Fischer}, \citenamefont {Friemann}, \citenamefont
  {Enkavi}, \citenamefont {Tajkhorshid},\ and\ \citenamefont
  {Neutze}}]{ErikssonFischerFriemannEnkaviTajkhorshidNeutze2013}%
  \BibitemOpen
  \bibfield  {author} {\bibinfo {author} {\bibfnamefont {U.~K.}\ \bibnamefont
  {Eriksson}}, \bibinfo {author} {\bibfnamefont {G.}~\bibnamefont {Fischer}},
  \bibinfo {author} {\bibfnamefont {R.}~\bibnamefont {Friemann}}, \bibinfo
  {author} {\bibfnamefont {G.}~\bibnamefont {Enkavi}}, \bibinfo {author}
  {\bibfnamefont {E.}~\bibnamefont {Tajkhorshid}}, \ and\ \bibinfo {author}
  {\bibfnamefont {R.}~\bibnamefont {Neutze}},\ }\href@noop {} {\bibfield
  {journal} {\bibinfo  {journal} {Science}\ }\textbf {\bibinfo {volume}
  {340}},\ \bibinfo {pages} {1346} (\bibinfo {year} {2013})}\BibitemShut
  {NoStop}%
\bibitem [{\citenamefont {Horner}\ \emph {et~al.}(2015)\citenamefont {Horner},
  \citenamefont {Zocher}, \citenamefont {Preiner}, \citenamefont {Ollinger},
  \citenamefont {Siligan}, \citenamefont {Akimov},\ and\ \citenamefont
  {Pohl}}]{HornerZocherPreinerOllingerSiliganAkimovPohl2015}%
  \BibitemOpen
  \bibfield  {author} {\bibinfo {author} {\bibfnamefont {A.}~\bibnamefont
  {Horner}}, \bibinfo {author} {\bibfnamefont {F.}~\bibnamefont {Zocher}},
  \bibinfo {author} {\bibfnamefont {J.}~\bibnamefont {Preiner}}, \bibinfo
  {author} {\bibfnamefont {N.}~\bibnamefont {Ollinger}}, \bibinfo {author}
  {\bibfnamefont {C.}~\bibnamefont {Siligan}}, \bibinfo {author} {\bibfnamefont
  {S.~A.}\ \bibnamefont {Akimov}}, \ and\ \bibinfo {author} {\bibfnamefont
  {P.}~\bibnamefont {Pohl}},\ }\href@noop {} {\bibfield  {journal} {\bibinfo
  {journal} {Science Adv.}\ }\textbf {\bibinfo {volume} {1}},\ \bibinfo {pages}
  {e1400083} (\bibinfo {year} {2015})}\BibitemShut {NoStop}%
\bibitem [{\citenamefont {Shen}\ \emph {et~al.}(2015)\citenamefont {Shen},
  \citenamefont {Si}, \citenamefont {Erbakan}, \citenamefont {Decker},
  \citenamefont {De~Zorzi}, \citenamefont {Saboe}, \citenamefont {Kang},
  \citenamefont {Majd}, \citenamefont {Butler}, \citenamefont {Walz},
  \citenamefont {Aksimentiev}, \citenamefont {li~Hou},\ and\ \citenamefont
  {Kumar}}]{ShenSiErbakanDeckerDeSaboeKangMajdButlerWalzAksimentievHouKumar2015}%
  \BibitemOpen
  \bibfield  {author} {\bibinfo {author} {\bibfnamefont {Y.-x.}\ \bibnamefont
  {Shen}}, \bibinfo {author} {\bibfnamefont {W.}~\bibnamefont {Si}}, \bibinfo
  {author} {\bibfnamefont {M.}~\bibnamefont {Erbakan}}, \bibinfo {author}
  {\bibfnamefont {K.}~\bibnamefont {Decker}}, \bibinfo {author} {\bibfnamefont
  {R.}~\bibnamefont {De~Zorzi}}, \bibinfo {author} {\bibfnamefont {P.~O.}\
  \bibnamefont {Saboe}}, \bibinfo {author} {\bibfnamefont {Y.~J.}\ \bibnamefont
  {Kang}}, \bibinfo {author} {\bibfnamefont {S.}~\bibnamefont {Majd}}, \bibinfo
  {author} {\bibfnamefont {P.~J.}\ \bibnamefont {Butler}}, \bibinfo {author}
  {\bibfnamefont {T.}~\bibnamefont {Walz}}, \bibinfo {author} {\bibfnamefont
  {A.}~\bibnamefont {Aksimentiev}}, \bibinfo {author} {\bibfnamefont
  {J.}~\bibnamefont {li~Hou}}, \ and\ \bibinfo {author} {\bibfnamefont
  {M.}~\bibnamefont {Kumar}},\ }\href@noop {} {\bibfield  {journal} {\bibinfo
  {journal} {Proc. Natl. Acad. Sci. USA}\ }\textbf {\bibinfo {volume} {112}},\
  \bibinfo {pages} {9810} (\bibinfo {year} {2015})}\BibitemShut {NoStop}%
\bibitem [{\citenamefont {Licsandru}\ \emph {et~al.}(2016)\citenamefont
  {Licsandru}, \citenamefont {Kocsis}, \citenamefont {Shen}, \citenamefont
  {Murail}, \citenamefont {Legrand}, \citenamefont {Van Der~Lee}, \citenamefont
  {Tsai}, \citenamefont {Baaden}, \citenamefont {Kumar},\ and\ \citenamefont
  {Barboiu}}]{LicsandruKocsisShenMurailLegrandVanDerLeeTsaiBaadenKumarBarboiu2016}%
  \BibitemOpen
  \bibfield  {author} {\bibinfo {author} {\bibfnamefont {E.}~\bibnamefont
  {Licsandru}}, \bibinfo {author} {\bibfnamefont {I.}~\bibnamefont {Kocsis}},
  \bibinfo {author} {\bibfnamefont {Y.-x.}\ \bibnamefont {Shen}}, \bibinfo
  {author} {\bibfnamefont {S.}~\bibnamefont {Murail}}, \bibinfo {author}
  {\bibfnamefont {Y.-M.}\ \bibnamefont {Legrand}}, \bibinfo {author}
  {\bibfnamefont {A.}~\bibnamefont {Van Der~Lee}}, \bibinfo {author}
  {\bibfnamefont {D.}~\bibnamefont {Tsai}}, \bibinfo {author} {\bibfnamefont
  {M.}~\bibnamefont {Baaden}}, \bibinfo {author} {\bibfnamefont
  {M.}~\bibnamefont {Kumar}}, \ and\ \bibinfo {author} {\bibfnamefont
  {M.}~\bibnamefont {Barboiu}},\ }\href@noop {} {\bibfield  {journal} {\bibinfo
   {journal} {J. Am. Chem. Soc.}\ }\textbf {\bibinfo {volume} {138}},\ \bibinfo
  {pages} {5403} (\bibinfo {year} {2016})}\BibitemShut {NoStop}%
\bibitem [{\citenamefont {Shen}\ \emph {et~al.}(2018)\citenamefont {Shen},
  \citenamefont {Song}, \citenamefont {Barden}, \citenamefont {Ren},
  \citenamefont {Lang}, \citenamefont {Feroz}, \citenamefont {Henderson},
  \citenamefont {Saboe}, \citenamefont {Tsai}, \citenamefont {Yan},
  \citenamefont {Butler}, \citenamefont {Bazan}, \citenamefont {Phillip},
  \citenamefont {Hickey}, \citenamefont {Cremer}, \citenamefont {Vashisth},\
  and\ \citenamefont
  {Kumar}}]{ShenSongBardenRenLangFerozHendersonSaboeTsaiYanButlerBazanPhillipHickeyCremerVashisthKumar2018}%
  \BibitemOpen
  \bibfield  {author} {\bibinfo {author} {\bibfnamefont {Y.-x.}\ \bibnamefont
  {Shen}}, \bibinfo {author} {\bibfnamefont {W.}~\bibnamefont {Song}}, \bibinfo
  {author} {\bibfnamefont {D.~R.}\ \bibnamefont {Barden}}, \bibinfo {author}
  {\bibfnamefont {T.}~\bibnamefont {Ren}}, \bibinfo {author} {\bibfnamefont
  {C.}~\bibnamefont {Lang}}, \bibinfo {author} {\bibfnamefont {H.}~\bibnamefont
  {Feroz}}, \bibinfo {author} {\bibfnamefont {C.~B.}\ \bibnamefont
  {Henderson}}, \bibinfo {author} {\bibfnamefont {P.~O.}\ \bibnamefont
  {Saboe}}, \bibinfo {author} {\bibfnamefont {D.}~\bibnamefont {Tsai}},
  \bibinfo {author} {\bibfnamefont {H.}~\bibnamefont {Yan}}, \bibinfo {author}
  {\bibfnamefont {P.~J.}\ \bibnamefont {Butler}}, \bibinfo {author}
  {\bibfnamefont {G.~C.}\ \bibnamefont {Bazan}}, \bibinfo {author}
  {\bibfnamefont {W.~A.}\ \bibnamefont {Phillip}}, \bibinfo {author}
  {\bibfnamefont {R.~J.}\ \bibnamefont {Hickey}}, \bibinfo {author}
  {\bibfnamefont {P.~S.}\ \bibnamefont {Cremer}}, \bibinfo {author}
  {\bibfnamefont {H.}~\bibnamefont {Vashisth}}, \ and\ \bibinfo {author}
  {\bibfnamefont {M.}~\bibnamefont {Kumar}},\ }\href@noop {} {\bibfield
  {journal} {\bibinfo  {journal} {Nat. Commun.}\ }\textbf {\bibinfo {volume}
  {9}},\ \bibinfo {pages} {2294} (\bibinfo {year} {2018})}\BibitemShut
  {NoStop}%
\bibitem [{\citenamefont {Song}\ \emph {et~al.}(2020)\citenamefont {Song},
  \citenamefont {Joshi}, \citenamefont {Chowdhury}, \citenamefont {Najem},
  \citenamefont {Shen}, \citenamefont {Lang}, \citenamefont {Henderson},
  \citenamefont {Tu}, \citenamefont {Farell}, \citenamefont {Pitz},
  \citenamefont {Maranas}, \citenamefont {Cremer}, \citenamefont {Hickey},
  \citenamefont {Sarles}, \citenamefont {li~Hou}, \citenamefont {Aksimentiev},\
  and\ \citenamefont
  {Kumar}}]{SongJoshiChowdhuryNajemShenLangHendersonTuFarellPitzMaranasCremerHickeySarlesHouAksimentievKumar2020}%
  \BibitemOpen
  \bibfield  {author} {\bibinfo {author} {\bibfnamefont {W.}~\bibnamefont
  {Song}}, \bibinfo {author} {\bibfnamefont {H.}~\bibnamefont {Joshi}},
  \bibinfo {author} {\bibfnamefont {R.}~\bibnamefont {Chowdhury}}, \bibinfo
  {author} {\bibfnamefont {J.~S.}\ \bibnamefont {Najem}}, \bibinfo {author}
  {\bibfnamefont {Y.-x.}\ \bibnamefont {Shen}}, \bibinfo {author}
  {\bibfnamefont {C.}~\bibnamefont {Lang}}, \bibinfo {author} {\bibfnamefont
  {C.~B.}\ \bibnamefont {Henderson}}, \bibinfo {author} {\bibfnamefont {Y.-M.}\
  \bibnamefont {Tu}}, \bibinfo {author} {\bibfnamefont {M.}~\bibnamefont
  {Farell}}, \bibinfo {author} {\bibfnamefont {M.~E.}\ \bibnamefont {Pitz}},
  \bibinfo {author} {\bibfnamefont {C.~D.}\ \bibnamefont {Maranas}}, \bibinfo
  {author} {\bibfnamefont {P.~S.}\ \bibnamefont {Cremer}}, \bibinfo {author}
  {\bibfnamefont {R.~J.}\ \bibnamefont {Hickey}}, \bibinfo {author}
  {\bibfnamefont {S.~A.}\ \bibnamefont {Sarles}}, \bibinfo {author}
  {\bibfnamefont {J.}~\bibnamefont {li~Hou}}, \bibinfo {author} {\bibfnamefont
  {A.}~\bibnamefont {Aksimentiev}}, \ and\ \bibinfo {author} {\bibfnamefont
  {M.}~\bibnamefont {Kumar}},\ }\href@noop {} {\bibfield  {journal} {\bibinfo
  {journal} {Nat. Nanotech.}\ }\textbf {\bibinfo {volume} {15}},\ \bibinfo
  {pages} {73} (\bibinfo {year} {2020})}\BibitemShut {NoStop}%
\bibitem [{\citenamefont {Shen}\ \emph {et~al.}(2020)\citenamefont {Shen},
  \citenamefont {Ye}, \citenamefont {Romanies}, \citenamefont {Roy},
  \citenamefont {Chen}, \citenamefont {Ren}, \citenamefont {Liu},\ and\
  \citenamefont {Zeng}}]{ShenYeRomaniesRoyChenRenLiuZeng2020}%
  \BibitemOpen
  \bibfield  {author} {\bibinfo {author} {\bibfnamefont {J.}~\bibnamefont
  {Shen}}, \bibinfo {author} {\bibfnamefont {R.}~\bibnamefont {Ye}}, \bibinfo
  {author} {\bibfnamefont {A.}~\bibnamefont {Romanies}}, \bibinfo {author}
  {\bibfnamefont {A.}~\bibnamefont {Roy}}, \bibinfo {author} {\bibfnamefont
  {F.}~\bibnamefont {Chen}}, \bibinfo {author} {\bibfnamefont {C.}~\bibnamefont
  {Ren}}, \bibinfo {author} {\bibfnamefont {Z.}~\bibnamefont {Liu}}, \ and\
  \bibinfo {author} {\bibfnamefont {H.}~\bibnamefont {Zeng}},\ }\href@noop {}
  {\bibfield  {journal} {\bibinfo  {journal} {J. Am. Chem. Soc.}\ }\textbf
  {\bibinfo {volume} {142}},\ \bibinfo {pages} {10050} (\bibinfo {year}
  {2020})}\BibitemShut {NoStop}%
\bibitem [{\citenamefont {Porter}\ \emph {et~al.}(2020)\citenamefont {Porter},
  \citenamefont {Werber}, \citenamefont {Zhong}, \citenamefont {Wilson},\ and\
  \citenamefont {Elimelech}}]{PorterWerberZhongWilsonElimelech2020}%
  \BibitemOpen
  \bibfield  {author} {\bibinfo {author} {\bibfnamefont {C.~J.}\ \bibnamefont
  {Porter}}, \bibinfo {author} {\bibfnamefont {J.~R.}\ \bibnamefont {Werber}},
  \bibinfo {author} {\bibfnamefont {M.}~\bibnamefont {Zhong}}, \bibinfo
  {author} {\bibfnamefont {C.~J.}\ \bibnamefont {Wilson}}, \ and\ \bibinfo
  {author} {\bibfnamefont {M.}~\bibnamefont {Elimelech}},\ }\href@noop {}
  {\bibfield  {journal} {\bibinfo  {journal} {ACS Nano}\ }\textbf {\bibinfo
  {volume} {14}},\ \bibinfo {pages} {10894} (\bibinfo {year}
  {2020})}\BibitemShut {NoStop}%
\bibitem [{\citenamefont {Roy}\ \emph {et~al.}(2021)\citenamefont {Roy},
  \citenamefont {Shen}, \citenamefont {Joshi}, \citenamefont {Song},
  \citenamefont {Tu}, \citenamefont {Chowdhury}, \citenamefont {Ye},
  \citenamefont {Li}, \citenamefont {Ren}, \citenamefont {Kumar}, \citenamefont
  {Aksimentiev},\ and\ \citenamefont
  {Zeng}}]{RoyShenJoshiSongTuChowdhuryYeLiRenKumarAksimentievZeng2021}%
  \BibitemOpen
  \bibfield  {author} {\bibinfo {author} {\bibfnamefont {A.}~\bibnamefont
  {Roy}}, \bibinfo {author} {\bibfnamefont {J.}~\bibnamefont {Shen}}, \bibinfo
  {author} {\bibfnamefont {H.}~\bibnamefont {Joshi}}, \bibinfo {author}
  {\bibfnamefont {W.}~\bibnamefont {Song}}, \bibinfo {author} {\bibfnamefont
  {Y.-M.}\ \bibnamefont {Tu}}, \bibinfo {author} {\bibfnamefont
  {R.}~\bibnamefont {Chowdhury}}, \bibinfo {author} {\bibfnamefont
  {R.}~\bibnamefont {Ye}}, \bibinfo {author} {\bibfnamefont {N.}~\bibnamefont
  {Li}}, \bibinfo {author} {\bibfnamefont {C.}~\bibnamefont {Ren}}, \bibinfo
  {author} {\bibfnamefont {M.}~\bibnamefont {Kumar}}, \bibinfo {author}
  {\bibfnamefont {A.}~\bibnamefont {Aksimentiev}}, \ and\ \bibinfo {author}
  {\bibfnamefont {H.}~\bibnamefont {Zeng}},\ }\href@noop {} {\bibfield
  {journal} {\bibinfo  {journal} {Nat. Nanotech.}\ }\textbf {\bibinfo {volume}
  {16}},\ \bibinfo {pages} {911} (\bibinfo {year} {2021})}\BibitemShut
  {NoStop}%
\bibitem [{\citenamefont {Shen}\ \emph
  {et~al.}(2021{\natexlab{b}})\citenamefont {Shen}, \citenamefont {Fei},
  \citenamefont {Zhong}, \citenamefont {Fan}, \citenamefont {Sun},
  \citenamefont {Hu}, \citenamefont {Gong}, \citenamefont {Czajkowsky},\ and\
  \citenamefont {Shao}}]{ShenFeiZhongFanSunHuGongCzajkowskyShao2021}%
  \BibitemOpen
  \bibfield  {author} {\bibinfo {author} {\bibfnamefont {Y.}~\bibnamefont
  {Shen}}, \bibinfo {author} {\bibfnamefont {F.}~\bibnamefont {Fei}}, \bibinfo
  {author} {\bibfnamefont {Y.}~\bibnamefont {Zhong}}, \bibinfo {author}
  {\bibfnamefont {C.}~\bibnamefont {Fan}}, \bibinfo {author} {\bibfnamefont
  {J.}~\bibnamefont {Sun}}, \bibinfo {author} {\bibfnamefont {J.}~\bibnamefont
  {Hu}}, \bibinfo {author} {\bibfnamefont {B.}~\bibnamefont {Gong}}, \bibinfo
  {author} {\bibfnamefont {D.~M.}\ \bibnamefont {Czajkowsky}}, \ and\ \bibinfo
  {author} {\bibfnamefont {Z.}~\bibnamefont {Shao}},\ }\href@noop {} {\bibfield
   {journal} {\bibinfo  {journal} {ACS Cent. Sci.}\ }\textbf {\bibinfo {volume}
  {7}},\ \bibinfo {pages} {2092} (\bibinfo {year}
  {2021}{\natexlab{b}})}\BibitemShut {NoStop}%
\bibitem [{\citenamefont {Tunuguntla}\ \emph {et~al.}(2017)\citenamefont
  {Tunuguntla}, \citenamefont {Henley}, \citenamefont {Yao}, \citenamefont
  {Pham}, \citenamefont {Wanunu},\ and\ \citenamefont
  {Noy}}]{TunuguntlaHenleyYaoPhamWanunuNoy2017}%
  \BibitemOpen
  \bibfield  {author} {\bibinfo {author} {\bibfnamefont {R.~H.}\ \bibnamefont
  {Tunuguntla}}, \bibinfo {author} {\bibfnamefont {R.~Y.}\ \bibnamefont
  {Henley}}, \bibinfo {author} {\bibfnamefont {Y.-C.}\ \bibnamefont {Yao}},
  \bibinfo {author} {\bibfnamefont {T.~A.}\ \bibnamefont {Pham}}, \bibinfo
  {author} {\bibfnamefont {M.}~\bibnamefont {Wanunu}}, \ and\ \bibinfo {author}
  {\bibfnamefont {A.}~\bibnamefont {Noy}},\ }\href@noop {} {\bibfield
  {journal} {\bibinfo  {journal} {Science}\ }\textbf {\bibinfo {volume}
  {357}},\ \bibinfo {pages} {792} (\bibinfo {year} {2017})}\BibitemShut
  {NoStop}%
\bibitem [{\citenamefont {Li}\ \emph {et~al.}(2020)\citenamefont {Li},
  \citenamefont {Li}, \citenamefont {Aydin}, \citenamefont {Quan},
  \citenamefont {Chen}, \citenamefont {Yao}, \citenamefont {Zhan},
  \citenamefont {Chen}, \citenamefont {Pham},\ and\ \citenamefont
  {Noy}}]{LiLiAydinQuanChenYaoZhanChenPhamNoy2020}%
  \BibitemOpen
  \bibfield  {author} {\bibinfo {author} {\bibfnamefont {Y.}~\bibnamefont
  {Li}}, \bibinfo {author} {\bibfnamefont {Z.}~\bibnamefont {Li}}, \bibinfo
  {author} {\bibfnamefont {F.}~\bibnamefont {Aydin}}, \bibinfo {author}
  {\bibfnamefont {J.}~\bibnamefont {Quan}}, \bibinfo {author} {\bibfnamefont
  {X.}~\bibnamefont {Chen}}, \bibinfo {author} {\bibfnamefont {Y.-C.}\
  \bibnamefont {Yao}}, \bibinfo {author} {\bibfnamefont {C.}~\bibnamefont
  {Zhan}}, \bibinfo {author} {\bibfnamefont {Y.}~\bibnamefont {Chen}}, \bibinfo
  {author} {\bibfnamefont {T.~A.}\ \bibnamefont {Pham}}, \ and\ \bibinfo
  {author} {\bibfnamefont {A.}~\bibnamefont {Noy}},\ }\href@noop {} {\bibfield
  {journal} {\bibinfo  {journal} {Science Adv.}\ }\textbf {\bibinfo {volume}
  {6}},\ \bibinfo {pages} {eaba9966} (\bibinfo {year} {2020})}\BibitemShut
  {NoStop}%
\bibitem [{\citenamefont {Park}\ \emph {et~al.}(2017)\citenamefont {Park},
  \citenamefont {Kamcev}, \citenamefont {Robeson}, \citenamefont {Elimelech},\
  and\ \citenamefont {Freeman}}]{ParkKamcevRobesonElimelechFreeman2017}%
  \BibitemOpen
  \bibfield  {author} {\bibinfo {author} {\bibfnamefont {H.~B.}\ \bibnamefont
  {Park}}, \bibinfo {author} {\bibfnamefont {J.}~\bibnamefont {Kamcev}},
  \bibinfo {author} {\bibfnamefont {L.~M.}\ \bibnamefont {Robeson}}, \bibinfo
  {author} {\bibfnamefont {M.}~\bibnamefont {Elimelech}}, \ and\ \bibinfo
  {author} {\bibfnamefont {B.~D.}\ \bibnamefont {Freeman}},\ }\href@noop {}
  {\bibfield  {journal} {\bibinfo  {journal} {Science}\ }\textbf {\bibinfo
  {volume} {356}},\ \bibinfo {pages} {eaab0530} (\bibinfo {year}
  {2017})}\BibitemShut {NoStop}%
\bibitem [{\citenamefont {Hummer}\ \emph {et~al.}(2001)\citenamefont {Hummer},
  \citenamefont {Rasaiah},\ and\ \citenamefont
  {Noworyta}}]{HummerRasaiahNoworyta2001}%
  \BibitemOpen
  \bibfield  {author} {\bibinfo {author} {\bibfnamefont {G.}~\bibnamefont
  {Hummer}}, \bibinfo {author} {\bibfnamefont {J.~C.}\ \bibnamefont {Rasaiah}},
  \ and\ \bibinfo {author} {\bibfnamefont {J.~P.}\ \bibnamefont {Noworyta}},\
  }\href@noop {} {\bibfield  {journal} {\bibinfo  {journal} {Nature}\ }\textbf
  {\bibinfo {volume} {414}},\ \bibinfo {pages} {188} (\bibinfo {year}
  {2001})}\BibitemShut {NoStop}%
\bibitem [{\citenamefont {Sriraman}\ \emph {et~al.}(2005)\citenamefont
  {Sriraman}, \citenamefont {Kevrekidis},\ and\ \citenamefont
  {Hummer}}]{SriramanKevrekidisHummer2005}%
  \BibitemOpen
  \bibfield  {author} {\bibinfo {author} {\bibfnamefont {S.}~\bibnamefont
  {Sriraman}}, \bibinfo {author} {\bibfnamefont {I.~G.}\ \bibnamefont
  {Kevrekidis}}, \ and\ \bibinfo {author} {\bibfnamefont {G.}~\bibnamefont
  {Hummer}},\ }\href {\doibase 10.1103/PhysRevLett.95.130603} {\bibfield
  {journal} {\bibinfo  {journal} {Phys. Rev. Lett.}\ }\textbf {\bibinfo
  {volume} {95}},\ \bibinfo {pages} {130603} (\bibinfo {year}
  {2005})}\BibitemShut {NoStop}%
\bibitem [{\citenamefont {Corry}(2011)}]{Corry2011}%
  \BibitemOpen
  \bibfield  {author} {\bibinfo {author} {\bibfnamefont {B.}~\bibnamefont
  {Corry}},\ }\href@noop {} {\bibfield  {journal} {\bibinfo  {journal} {Energy
  Environ. Sci.}\ }\textbf {\bibinfo {volume} {4}},\ \bibinfo {pages} {751}
  (\bibinfo {year} {2011})}\BibitemShut {NoStop}%
\bibitem [{\citenamefont {Garc{\'\i}a-Fandi{\~n}o}\ and\ \citenamefont
  {Sansom}(2012)}]{Garc'ia-FandinoSansom2012}%
  \BibitemOpen
  \bibfield  {author} {\bibinfo {author} {\bibfnamefont {R.}~\bibnamefont
  {Garc{\'\i}a-Fandi{\~n}o}}\ and\ \bibinfo {author} {\bibfnamefont {M.~S.~P.}\
  \bibnamefont {Sansom}},\ }\href@noop {} {\bibfield  {journal} {\bibinfo
  {journal} {Proc. Natl. Acad. Sci. USA}\ }\textbf {\bibinfo {volume} {109}},\
  \bibinfo {pages} {6939} (\bibinfo {year} {2012})}\BibitemShut {NoStop}%
\bibitem [{\citenamefont {Winarto}\ \emph {et~al.}(2015)\citenamefont
  {Winarto}, \citenamefont {Takaiwa}, \citenamefont {Yamamoto},\ and\
  \citenamefont {Yasuoka}}]{WinartoTakaiwaYamamotoYasuoka2015a}%
  \BibitemOpen
  \bibfield  {author} {\bibinfo {author} {\bibnamefont {Winarto}}, \bibinfo
  {author} {\bibfnamefont {D.}~\bibnamefont {Takaiwa}}, \bibinfo {author}
  {\bibfnamefont {E.}~\bibnamefont {Yamamoto}}, \ and\ \bibinfo {author}
  {\bibfnamefont {K.}~\bibnamefont {Yasuoka}},\ }\href@noop {} {\bibfield
  {journal} {\bibinfo  {journal} {Nanoscale}\ }\textbf {\bibinfo {volume}
  {7}},\ \bibinfo {pages} {12659} (\bibinfo {year} {2015})}\BibitemShut
  {NoStop}%
\bibitem [{\citenamefont {Zhou}\ and\ \citenamefont {Zhu}(2018)}]{ZhouZhu2018}%
  \BibitemOpen
  \bibfield  {author} {\bibinfo {author} {\bibfnamefont {X.}~\bibnamefont
  {Zhou}}\ and\ \bibinfo {author} {\bibfnamefont {F.}~\bibnamefont {Zhu}},\
  }\href {\doibase 10.1103/PhysRevE.98.032410} {\bibfield  {journal} {\bibinfo
  {journal} {Phys. Rev. E}\ }\textbf {\bibinfo {volume} {98}},\ \bibinfo
  {pages} {032410} (\bibinfo {year} {2018})}\BibitemShut {NoStop}%
\bibitem [{\citenamefont {Winarto}\ \emph {et~al.}(2019)\citenamefont
  {Winarto}, \citenamefont {Yamamoto},\ and\ \citenamefont
  {Yasuoka}}]{WinartoYamamotoYasuoka2019}%
  \BibitemOpen
  \bibfield  {author} {\bibinfo {author} {\bibnamefont {Winarto}}, \bibinfo
  {author} {\bibfnamefont {E.}~\bibnamefont {Yamamoto}}, \ and\ \bibinfo
  {author} {\bibfnamefont {K.}~\bibnamefont {Yasuoka}},\ }\href@noop {}
  {\bibfield  {journal} {\bibinfo  {journal} {Phys. Chem. Chem. Phys.}\
  }\textbf {\bibinfo {volume} {21}},\ \bibinfo {pages} {15431} (\bibinfo {year}
  {2019})}\BibitemShut {NoStop}%
\bibitem [{\citenamefont {Klesse}\ \emph {et~al.}(2020)\citenamefont {Klesse},
  \citenamefont {Tucker},\ and\ \citenamefont
  {Sansom}}]{KlesseTuckerSansom2020}%
  \BibitemOpen
  \bibfield  {author} {\bibinfo {author} {\bibfnamefont {G.}~\bibnamefont
  {Klesse}}, \bibinfo {author} {\bibfnamefont {S.~J.}\ \bibnamefont {Tucker}},
  \ and\ \bibinfo {author} {\bibfnamefont {M.~S.~P.}\ \bibnamefont {Sansom}},\
  }\href@noop {} {\bibfield  {journal} {\bibinfo  {journal} {ACS Nano}\
  }\textbf {\bibinfo {volume} {14}},\ \bibinfo {pages} {10480} (\bibinfo {year}
  {2020})}\BibitemShut {NoStop}%
\bibitem [{\citenamefont {Arai}\ \emph {et~al.}(2021)\citenamefont {Arai},
  \citenamefont {Koishi},\ and\ \citenamefont
  {Ebisuzaki}}]{AraiKoishiEbisuzaki2021}%
  \BibitemOpen
  \bibfield  {author} {\bibinfo {author} {\bibfnamefont {N.}~\bibnamefont
  {Arai}}, \bibinfo {author} {\bibfnamefont {T.}~\bibnamefont {Koishi}}, \ and\
  \bibinfo {author} {\bibfnamefont {T.}~\bibnamefont {Ebisuzaki}},\ }\href@noop
  {} {\bibfield  {journal} {\bibinfo  {journal} {ACS Nano}\ }\textbf {\bibinfo
  {volume} {15}},\ \bibinfo {pages} {2481} (\bibinfo {year}
  {2021})}\BibitemShut {NoStop}%
\bibitem [{\citenamefont {Lu}\ \emph {et~al.}(2012)\citenamefont {Lu},
  \citenamefont {Nie}, \citenamefont {Wu}, \citenamefont {Zhou}, \citenamefont
  {Kou}, \citenamefont {Xu},\ and\ \citenamefont
  {Liu}}]{LuNieWuZhouKouXuLiu2012}%
  \BibitemOpen
  \bibfield  {author} {\bibinfo {author} {\bibfnamefont {H.}~\bibnamefont
  {Lu}}, \bibinfo {author} {\bibfnamefont {X.}~\bibnamefont {Nie}}, \bibinfo
  {author} {\bibfnamefont {F.}~\bibnamefont {Wu}}, \bibinfo {author}
  {\bibfnamefont {X.}~\bibnamefont {Zhou}}, \bibinfo {author} {\bibfnamefont
  {J.}~\bibnamefont {Kou}}, \bibinfo {author} {\bibfnamefont {Y.}~\bibnamefont
  {Xu}}, \ and\ \bibinfo {author} {\bibfnamefont {Y.}~\bibnamefont {Liu}},\
  }\href@noop {} {\bibfield  {journal} {\bibinfo  {journal} {J. Chem. Phys.}\
  }\textbf {\bibinfo {volume} {136}},\ \bibinfo {pages} {174511} (\bibinfo
  {year} {2012})}\BibitemShut {NoStop}%
\bibitem [{\citenamefont {Kou}\ \emph {et~al.}(2012)\citenamefont {Kou},
  \citenamefont {Zhou}, \citenamefont {Lu}, \citenamefont {Xu}, \citenamefont
  {Wu},\ and\ \citenamefont {Fan}}]{KouZhouLuXuWuFan2012}%
  \BibitemOpen
  \bibfield  {author} {\bibinfo {author} {\bibfnamefont {J.}~\bibnamefont
  {Kou}}, \bibinfo {author} {\bibfnamefont {X.}~\bibnamefont {Zhou}}, \bibinfo
  {author} {\bibfnamefont {H.}~\bibnamefont {Lu}}, \bibinfo {author}
  {\bibfnamefont {Y.}~\bibnamefont {Xu}}, \bibinfo {author} {\bibfnamefont
  {F.}~\bibnamefont {Wu}}, \ and\ \bibinfo {author} {\bibfnamefont
  {J.}~\bibnamefont {Fan}},\ }\href@noop {} {\bibfield  {journal} {\bibinfo
  {journal} {Soft Matter}\ }\textbf {\bibinfo {volume} {8}},\ \bibinfo {pages}
  {12111} (\bibinfo {year} {2012})}\BibitemShut {NoStop}%
\bibitem [{\citenamefont {Zhou}\ \emph {et~al.}(2013)\citenamefont {Zhou},
  \citenamefont {Wu}, \citenamefont {Kou}, \citenamefont {Nie}, \citenamefont
  {Yang},\ and\ \citenamefont {Lu}}]{ZhouWuKouNieYangLu2013}%
  \BibitemOpen
  \bibfield  {author} {\bibinfo {author} {\bibfnamefont {X.}~\bibnamefont
  {Zhou}}, \bibinfo {author} {\bibfnamefont {F.-M.}\ \bibnamefont {Wu}},
  \bibinfo {author} {\bibfnamefont {J.}~\bibnamefont {Kou}}, \bibinfo {author}
  {\bibfnamefont {X.}~\bibnamefont {Nie}}, \bibinfo {author} {\bibfnamefont
  {L.}~\bibnamefont {Yang}}, \ and\ \bibinfo {author} {\bibfnamefont
  {H.}~\bibnamefont {Lu}},\ }\href@noop {} {\bibfield  {journal} {\bibinfo
  {journal} {J. Phys. Chem. B}\ }\textbf {\bibinfo {volume} {117}},\ \bibinfo
  {pages} {11681} (\bibinfo {year} {2013})}\BibitemShut {NoStop}%
\bibitem [{\citenamefont {Liu}\ \emph {et~al.}(2016)\citenamefont {Liu},
  \citenamefont {Wang}, \citenamefont {Xie}, \citenamefont {Ju},\ and\
  \citenamefont {Chu}}]{LiuWangXieJuChu2016}%
  \BibitemOpen
  \bibfield  {author} {\bibinfo {author} {\bibfnamefont {Z.}~\bibnamefont
  {Liu}}, \bibinfo {author} {\bibfnamefont {W.}~\bibnamefont {Wang}}, \bibinfo
  {author} {\bibfnamefont {R.}~\bibnamefont {Xie}}, \bibinfo {author}
  {\bibfnamefont {X.-J.}\ \bibnamefont {Ju}}, \ and\ \bibinfo {author}
  {\bibfnamefont {L.-Y.}\ \bibnamefont {Chu}},\ }\href@noop {} {\bibfield
  {journal} {\bibinfo  {journal} {Chem. Soc. Rev.}\ }\textbf {\bibinfo {volume}
  {45}},\ \bibinfo {pages} {460} (\bibinfo {year} {2016})}\BibitemShut
  {NoStop}%
\bibitem [{\citenamefont {Zhu}\ \emph {et~al.}(2019)\citenamefont {Zhu},
  \citenamefont {Wang}, \citenamefont {Tian},\ and\ \citenamefont
  {Jiang}}]{ZhuWangTianJiang2019}%
  \BibitemOpen
  \bibfield  {author} {\bibinfo {author} {\bibfnamefont {Z.}~\bibnamefont
  {Zhu}}, \bibinfo {author} {\bibfnamefont {D.}~\bibnamefont {Wang}}, \bibinfo
  {author} {\bibfnamefont {Y.}~\bibnamefont {Tian}}, \ and\ \bibinfo {author}
  {\bibfnamefont {L.}~\bibnamefont {Jiang}},\ }\href@noop {} {\bibfield
  {journal} {\bibinfo  {journal} {J. Am. Chem. Soc.}\ }\textbf {\bibinfo
  {volume} {141}},\ \bibinfo {pages} {8658} (\bibinfo {year}
  {2019})}\BibitemShut {NoStop}%
\bibitem [{\citenamefont {Rinne}\ \emph {et~al.}(2012)\citenamefont {Rinne},
  \citenamefont {Gekle}, \citenamefont {Bonthuis},\ and\ \citenamefont
  {Netz}}]{RinneGekleBonthuisNetz2012}%
  \BibitemOpen
  \bibfield  {author} {\bibinfo {author} {\bibfnamefont {K.~F.}\ \bibnamefont
  {Rinne}}, \bibinfo {author} {\bibfnamefont {S.}~\bibnamefont {Gekle}},
  \bibinfo {author} {\bibfnamefont {D.~J.}\ \bibnamefont {Bonthuis}}, \ and\
  \bibinfo {author} {\bibfnamefont {R.~R.}\ \bibnamefont {Netz}},\ }\href@noop
  {} {\bibfield  {journal} {\bibinfo  {journal} {Nano Lett.}\ }\textbf
  {\bibinfo {volume} {12}},\ \bibinfo {pages} {1780} (\bibinfo {year}
  {2012})}\BibitemShut {NoStop}%
\bibitem [{\citenamefont {Xiao}\ \emph {et~al.}(2016)\citenamefont {Xiao},
  \citenamefont {Zhou}, \citenamefont {Kong}, \citenamefont {Xie},
  \citenamefont {Li}, \citenamefont {Zhang}, \citenamefont {Wen},\ and\
  \citenamefont {Jiang}}]{XiaoZhouKongXieLiZhangWenJiang2016}%
  \BibitemOpen
  \bibfield  {author} {\bibinfo {author} {\bibfnamefont {K.}~\bibnamefont
  {Xiao}}, \bibinfo {author} {\bibfnamefont {Y.}~\bibnamefont {Zhou}}, \bibinfo
  {author} {\bibfnamefont {X.-Y.}\ \bibnamefont {Kong}}, \bibinfo {author}
  {\bibfnamefont {G.}~\bibnamefont {Xie}}, \bibinfo {author} {\bibfnamefont
  {P.}~\bibnamefont {Li}}, \bibinfo {author} {\bibfnamefont {Z.}~\bibnamefont
  {Zhang}}, \bibinfo {author} {\bibfnamefont {L.}~\bibnamefont {Wen}}, \ and\
  \bibinfo {author} {\bibfnamefont {L.}~\bibnamefont {Jiang}},\ }\href@noop {}
  {\bibfield  {journal} {\bibinfo  {journal} {ACS Nano}\ }\textbf {\bibinfo
  {volume} {10}},\ \bibinfo {pages} {9703} (\bibinfo {year}
  {2016})}\BibitemShut {NoStop}%
\bibitem [{\citenamefont {Cai}\ \emph {et~al.}(2021)\citenamefont {Cai},
  \citenamefont {Ma}, \citenamefont {Hao}, \citenamefont {Sun}, \citenamefont
  {Guo}, \citenamefont {Xu}, \citenamefont {Xu},\ and\ \citenamefont
  {Kuang}}]{CaiMaHaoSunGuoXuXuKuang2021}%
  \BibitemOpen
  \bibfield  {author} {\bibinfo {author} {\bibfnamefont {J.}~\bibnamefont
  {Cai}}, \bibinfo {author} {\bibfnamefont {W.}~\bibnamefont {Ma}}, \bibinfo
  {author} {\bibfnamefont {C.}~\bibnamefont {Hao}}, \bibinfo {author}
  {\bibfnamefont {M.}~\bibnamefont {Sun}}, \bibinfo {author} {\bibfnamefont
  {J.}~\bibnamefont {Guo}}, \bibinfo {author} {\bibfnamefont {L.}~\bibnamefont
  {Xu}}, \bibinfo {author} {\bibfnamefont {C.}~\bibnamefont {Xu}}, \ and\
  \bibinfo {author} {\bibfnamefont {H.}~\bibnamefont {Kuang}},\ }\href@noop {}
  {\bibfield  {journal} {\bibinfo  {journal} {Chem}\ }\textbf {\bibinfo
  {volume} {7}},\ \bibinfo {pages} {1802} (\bibinfo {year} {2021})}\BibitemShut
  {NoStop}%
\bibitem [{\citenamefont {Li}\ \emph {et~al.}(2021)\citenamefont {Li},
  \citenamefont {Chang}, \citenamefont {Zhu}, \citenamefont {Sun},\ and\
  \citenamefont {Fan}}]{LiChangZhuSunFan2021}%
  \BibitemOpen
  \bibfield  {author} {\bibinfo {author} {\bibfnamefont {Y.}~\bibnamefont
  {Li}}, \bibinfo {author} {\bibfnamefont {C.}~\bibnamefont {Chang}}, \bibinfo
  {author} {\bibfnamefont {Z.}~\bibnamefont {Zhu}}, \bibinfo {author}
  {\bibfnamefont {L.}~\bibnamefont {Sun}}, \ and\ \bibinfo {author}
  {\bibfnamefont {C.}~\bibnamefont {Fan}},\ }\href@noop {} {\bibfield
  {journal} {\bibinfo  {journal} {J. Am. Chem. Soc.}\ }\textbf {\bibinfo
  {volume} {143}},\ \bibinfo {pages} {4311} (\bibinfo {year}
  {2021})}\BibitemShut {NoStop}%
\bibitem [{\citenamefont {Zhang}\ \emph {et~al.}(2013)\citenamefont {Zhang},
  \citenamefont {Hou}, \citenamefont {Zeng}, \citenamefont {Yang},
  \citenamefont {Li}, \citenamefont {Yan}, \citenamefont {Tian},\ and\
  \citenamefont {Jiang}}]{ZhangHouZengYangLiYanTianJiang2013}%
  \BibitemOpen
  \bibfield  {author} {\bibinfo {author} {\bibfnamefont {H.}~\bibnamefont
  {Zhang}}, \bibinfo {author} {\bibfnamefont {X.}~\bibnamefont {Hou}}, \bibinfo
  {author} {\bibfnamefont {L.}~\bibnamefont {Zeng}}, \bibinfo {author}
  {\bibfnamefont {F.}~\bibnamefont {Yang}}, \bibinfo {author} {\bibfnamefont
  {L.}~\bibnamefont {Li}}, \bibinfo {author} {\bibfnamefont {D.}~\bibnamefont
  {Yan}}, \bibinfo {author} {\bibfnamefont {Y.}~\bibnamefont {Tian}}, \ and\
  \bibinfo {author} {\bibfnamefont {L.}~\bibnamefont {Jiang}},\ }\href@noop {}
  {\bibfield  {journal} {\bibinfo  {journal} {J. Am. Chem. Soc.}\ }\textbf
  {\bibinfo {volume} {135}},\ \bibinfo {pages} {16102} (\bibinfo {year}
  {2013})}\BibitemShut {NoStop}%
\bibitem [{\citenamefont {Rao}\ \emph {et~al.}(2019)\citenamefont {Rao},
  \citenamefont {Klesse}, \citenamefont {Stansfeld}, \citenamefont {Tucker},\
  and\ \citenamefont {Sansom}}]{RaoKlesseStansfeldTuckerSansom2019}%
  \BibitemOpen
  \bibfield  {author} {\bibinfo {author} {\bibfnamefont {S.}~\bibnamefont
  {Rao}}, \bibinfo {author} {\bibfnamefont {G.}~\bibnamefont {Klesse}},
  \bibinfo {author} {\bibfnamefont {P.~J.}\ \bibnamefont {Stansfeld}}, \bibinfo
  {author} {\bibfnamefont {S.~J.}\ \bibnamefont {Tucker}}, \ and\ \bibinfo
  {author} {\bibfnamefont {M.~S.~P.}\ \bibnamefont {Sansom}},\ }\href@noop {}
  {\bibfield  {journal} {\bibinfo  {journal} {Proc. Natl. Acad. Sci. USA}\
  }\textbf {\bibinfo {volume} {116}},\ \bibinfo {pages} {13989} (\bibinfo
  {year} {2019})}\BibitemShut {NoStop}%
\bibitem [{\citenamefont {Beckstein}\ \emph {et~al.}(2001)\citenamefont
  {Beckstein}, \citenamefont {Biggin},\ and\ \citenamefont
  {Sansom}}]{BecksteinBigginSansom2001}%
  \BibitemOpen
  \bibfield  {author} {\bibinfo {author} {\bibfnamefont {O.}~\bibnamefont
  {Beckstein}}, \bibinfo {author} {\bibfnamefont {P.~C.}\ \bibnamefont
  {Biggin}}, \ and\ \bibinfo {author} {\bibfnamefont {M.~S.~P.}\ \bibnamefont
  {Sansom}},\ }\href@noop {} {\bibfield  {journal} {\bibinfo  {journal} {J.
  Phys. Chem. B}\ }\textbf {\bibinfo {volume} {105}},\ \bibinfo {pages} {12902}
  (\bibinfo {year} {2001})}\BibitemShut {NoStop}%
\bibitem [{\citenamefont {Beckstein}\ and\ \citenamefont
  {Sansom}(2003)}]{BecksteinSansom2003}%
  \BibitemOpen
  \bibfield  {author} {\bibinfo {author} {\bibfnamefont {O.}~\bibnamefont
  {Beckstein}}\ and\ \bibinfo {author} {\bibfnamefont {M.~S.~P.}\ \bibnamefont
  {Sansom}},\ }\href@noop {} {\bibfield  {journal} {\bibinfo  {journal} {Proc.
  Natl. Acad. Sci. USA}\ }\textbf {\bibinfo {volume} {100}},\ \bibinfo {pages}
  {7063} (\bibinfo {year} {2003})}\BibitemShut {NoStop}%
\bibitem [{\citenamefont {Kaptan}\ \emph {et~al.}(2015)\citenamefont {Kaptan},
  \citenamefont {Assentoft}, \citenamefont {Schneider}, \citenamefont {Fenton},
  \citenamefont {Deitmer}, \citenamefont {MacAulay},\ and\ \citenamefont
  {de~Groot}}]{KaptanAssentoftSchneiderFentonDeitmerMacAulayGroot2015}%
  \BibitemOpen
  \bibfield  {author} {\bibinfo {author} {\bibfnamefont {S.}~\bibnamefont
  {Kaptan}}, \bibinfo {author} {\bibfnamefont {M.}~\bibnamefont {Assentoft}},
  \bibinfo {author} {\bibfnamefont {H.~P.}\ \bibnamefont {Schneider}}, \bibinfo
  {author} {\bibfnamefont {R.~A.}\ \bibnamefont {Fenton}}, \bibinfo {author}
  {\bibfnamefont {J.~W.}\ \bibnamefont {Deitmer}}, \bibinfo {author}
  {\bibfnamefont {N.}~\bibnamefont {MacAulay}}, \ and\ \bibinfo {author}
  {\bibfnamefont {B.~L.}\ \bibnamefont {de~Groot}},\ }\href@noop {} {\bibfield
  {journal} {\bibinfo  {journal} {Structure}\ }\textbf {\bibinfo {volume}
  {23}},\ \bibinfo {pages} {2309} (\bibinfo {year} {2015})}\BibitemShut
  {NoStop}%
\bibitem [{\citenamefont {Lindahl}\ \emph {et~al.}(2018)\citenamefont
  {Lindahl}, \citenamefont {Gourdon}, \citenamefont {Andersson},\ and\
  \citenamefont {Hess}}]{LindahlGourdonAnderssonHess2018}%
  \BibitemOpen
  \bibfield  {author} {\bibinfo {author} {\bibfnamefont {V.}~\bibnamefont
  {Lindahl}}, \bibinfo {author} {\bibfnamefont {P.}~\bibnamefont {Gourdon}},
  \bibinfo {author} {\bibfnamefont {M.}~\bibnamefont {Andersson}}, \ and\
  \bibinfo {author} {\bibfnamefont {B.}~\bibnamefont {Hess}},\ }\href@noop {}
  {\bibfield  {journal} {\bibinfo  {journal} {Sci. Rep.}\ }\textbf {\bibinfo
  {volume} {8}},\ \bibinfo {pages} {2995} (\bibinfo {year} {2018})}\BibitemShut
  {NoStop}%
\bibitem [{\citenamefont {Yamamoto}\ \emph {et~al.}(2014)\citenamefont
  {Yamamoto}, \citenamefont {Akimoto}, \citenamefont {Hirano}, \citenamefont
  {Yasui},\ and\ \citenamefont
  {Yasuoka}}]{YamamotoAkimotoHiranoYasuiYasuoka2014}%
  \BibitemOpen
  \bibfield  {author} {\bibinfo {author} {\bibfnamefont {E.}~\bibnamefont
  {Yamamoto}}, \bibinfo {author} {\bibfnamefont {T.}~\bibnamefont {Akimoto}},
  \bibinfo {author} {\bibfnamefont {Y.}~\bibnamefont {Hirano}}, \bibinfo
  {author} {\bibfnamefont {M.}~\bibnamefont {Yasui}}, \ and\ \bibinfo {author}
  {\bibfnamefont {K.}~\bibnamefont {Yasuoka}},\ }\href {\doibase
  10.1103/PhysRevE.89.022718} {\bibfield  {journal} {\bibinfo  {journal} {Phys.
  Rev. E}\ }\textbf {\bibinfo {volume} {89}},\ \bibinfo {pages} {022718}
  (\bibinfo {year} {2014})}\BibitemShut {NoStop}%
\bibitem [{\citenamefont {M{\"a}nnikk{\"o}}\ \emph {et~al.}(2005)\citenamefont
  {M{\"a}nnikk{\"o}}, \citenamefont {Pandey}, \citenamefont {Larsson},\ and\
  \citenamefont {Elinder}}]{MaennikkoePandeyLarssonElinder2005}%
  \BibitemOpen
  \bibfield  {author} {\bibinfo {author} {\bibfnamefont {R.}~\bibnamefont
  {M{\"a}nnikk{\"o}}}, \bibinfo {author} {\bibfnamefont {S.}~\bibnamefont
  {Pandey}}, \bibinfo {author} {\bibfnamefont {H.~P.}\ \bibnamefont {Larsson}},
  \ and\ \bibinfo {author} {\bibfnamefont {F.}~\bibnamefont {Elinder}},\
  }\href@noop {} {\bibfield  {journal} {\bibinfo  {journal} {J. Gen. Physiol.}\
  }\textbf {\bibinfo {volume} {125}},\ \bibinfo {pages} {305} (\bibinfo {year}
  {2005})}\BibitemShut {NoStop}%
\bibitem [{\citenamefont {Tilegenova}\ \emph {et~al.}(2017)\citenamefont
  {Tilegenova}, \citenamefont {Cortes},\ and\ \citenamefont
  {Cuello}}]{TilegenovaCortesCuello2017}%
  \BibitemOpen
  \bibfield  {author} {\bibinfo {author} {\bibfnamefont {C.}~\bibnamefont
  {Tilegenova}}, \bibinfo {author} {\bibfnamefont {D.~M.}\ \bibnamefont
  {Cortes}}, \ and\ \bibinfo {author} {\bibfnamefont {L.~G.}\ \bibnamefont
  {Cuello}},\ }\href@noop {} {\bibfield  {journal} {\bibinfo  {journal} {Proc.
  Natl. Acad. Sci. USA}\ }\textbf {\bibinfo {volume} {114}},\ \bibinfo {pages}
  {3234} (\bibinfo {year} {2017})}\BibitemShut {NoStop}%
\bibitem [{\citenamefont {Villalba-Galea}(2017)}]{Villalba-Galea2017}%
  \BibitemOpen
  \bibfield  {author} {\bibinfo {author} {\bibfnamefont {C.~A.}\ \bibnamefont
  {Villalba-Galea}},\ }\href@noop {} {\bibfield  {journal} {\bibinfo  {journal}
  {Channels}\ }\textbf {\bibinfo {volume} {11}},\ \bibinfo {pages} {140}
  (\bibinfo {year} {2017})}\BibitemShut {NoStop}%
\bibitem [{\citenamefont {Goychuk}(2009)}]{Goychuk2009}%
  \BibitemOpen
  \bibfield  {author} {\bibinfo {author} {\bibfnamefont {I.}~\bibnamefont
  {Goychuk}},\ }\href {\doibase 10.1103/PhysRevE.80.046125} {\bibfield
  {journal} {\bibinfo  {journal} {Phys. Rev. E}\ }\textbf {\bibinfo {volume}
  {80}},\ \bibinfo {pages} {046125} (\bibinfo {year} {2009})}\BibitemShut
  {NoStop}%
\bibitem [{\citenamefont {Zeidel}\ \emph {et~al.}(1992)\citenamefont {Zeidel},
  \citenamefont {Ambudkar}, \citenamefont {Smith},\ and\ \citenamefont
  {Agre}}]{ZeidelAmbudkarSmithAgre1992}%
  \BibitemOpen
  \bibfield  {author} {\bibinfo {author} {\bibfnamefont {M.~L.}\ \bibnamefont
  {Zeidel}}, \bibinfo {author} {\bibfnamefont {S.~V.}\ \bibnamefont
  {Ambudkar}}, \bibinfo {author} {\bibfnamefont {B.~L.}\ \bibnamefont {Smith}},
  \ and\ \bibinfo {author} {\bibfnamefont {P.}~\bibnamefont {Agre}},\
  }\href@noop {} {\bibfield  {journal} {\bibinfo  {journal} {Biochemistry}\
  }\textbf {\bibinfo {volume} {31}},\ \bibinfo {pages} {7436} (\bibinfo {year}
  {1992})}\BibitemShut {NoStop}%
\bibitem [{\citenamefont {Borgnia}\ \emph {et~al.}(1999)\citenamefont
  {Borgnia}, \citenamefont {Nielsen}, \citenamefont {Engel},\ and\
  \citenamefont {Agre}}]{BorgniaNielsenEngelAgre1999}%
  \BibitemOpen
  \bibfield  {author} {\bibinfo {author} {\bibfnamefont {M.}~\bibnamefont
  {Borgnia}}, \bibinfo {author} {\bibfnamefont {S.}~\bibnamefont {Nielsen}},
  \bibinfo {author} {\bibfnamefont {A.}~\bibnamefont {Engel}}, \ and\ \bibinfo
  {author} {\bibfnamefont {P.}~\bibnamefont {Agre}},\ }\href@noop {} {\bibfield
   {journal} {\bibinfo  {journal} {Annu. Rev. Biochem.}\ }\textbf {\bibinfo
  {volume} {68}},\ \bibinfo {pages} {425} (\bibinfo {year} {1999})}\BibitemShut
  {NoStop}%
\bibitem [{\citenamefont {Hoogerbrugge}\ and\ \citenamefont
  {Koelman}(1992)}]{HoogerbruggeKoelman1992}%
  \BibitemOpen
  \bibfield  {author} {\bibinfo {author} {\bibfnamefont {P.~J.}\ \bibnamefont
  {Hoogerbrugge}}\ and\ \bibinfo {author} {\bibfnamefont {J.~M. V.~A.}\
  \bibnamefont {Koelman}},\ }\href@noop {} {\bibfield  {journal} {\bibinfo
  {journal} {Europhys. Lett.}\ }\textbf {\bibinfo {volume} {19}},\ \bibinfo
  {pages} {155} (\bibinfo {year} {1992})}\BibitemShut {NoStop}%
\bibitem [{\citenamefont {Groot}\ and\ \citenamefont
  {Warren}(1997)}]{GrootWarren1997}%
  \BibitemOpen
  \bibfield  {author} {\bibinfo {author} {\bibfnamefont {R.~D.}\ \bibnamefont
  {Groot}}\ and\ \bibinfo {author} {\bibfnamefont {P.~B.}\ \bibnamefont
  {Warren}},\ }\href@noop {} {\bibfield  {journal} {\bibinfo  {journal} {J.
  Chem. Phys.}\ }\textbf {\bibinfo {volume} {107}},\ \bibinfo {pages} {4423}
  (\bibinfo {year} {1997})}\BibitemShut {NoStop}%
\bibitem [{\citenamefont {Groot}\ and\ \citenamefont
  {Rabone}(2001)}]{GrootRabone2001}%
  \BibitemOpen
  \bibfield  {author} {\bibinfo {author} {\bibfnamefont {R.~D.}\ \bibnamefont
  {Groot}}\ and\ \bibinfo {author} {\bibfnamefont {K.~L.}\ \bibnamefont
  {Rabone}},\ }\href@noop {} {\bibfield  {journal} {\bibinfo  {journal}
  {Biophys. J.}\ }\textbf {\bibinfo {volume} {81}},\ \bibinfo {pages} {725}
  (\bibinfo {year} {2001})}\BibitemShut {NoStop}%
\bibitem [{\citenamefont {Arai}\ \emph {et~al.}(2010)\citenamefont {Arai},
  \citenamefont {Yasuoka}, \citenamefont {Koishi},\ and\ \citenamefont
  {Ebisuzaki}}]{AraiYasuokaKoishiEbisuzaki2010}%
  \BibitemOpen
  \bibfield  {author} {\bibinfo {author} {\bibfnamefont {N.}~\bibnamefont
  {Arai}}, \bibinfo {author} {\bibfnamefont {K.}~\bibnamefont {Yasuoka}},
  \bibinfo {author} {\bibfnamefont {T.}~\bibnamefont {Koishi}}, \ and\ \bibinfo
  {author} {\bibfnamefont {T.}~\bibnamefont {Ebisuzaki}},\ }\href@noop {}
  {\bibfield  {journal} {\bibinfo  {journal} {ACS Nano}\ }\textbf {\bibinfo
  {volume} {4}},\ \bibinfo {pages} {5905} (\bibinfo {year} {2010})}\BibitemShut
  {NoStop}%
\bibitem [{\citenamefont {Arai}\ \emph {et~al.}(2013)\citenamefont {Arai},
  \citenamefont {Yasuoka}, \citenamefont {Koishi}, \citenamefont {Ebisuzaki},\
  and\ \citenamefont {Zeng}}]{AraiYasuokaKoishiEbisuzakiZeng2013}%
  \BibitemOpen
  \bibfield  {author} {\bibinfo {author} {\bibfnamefont {N.}~\bibnamefont
  {Arai}}, \bibinfo {author} {\bibfnamefont {K.}~\bibnamefont {Yasuoka}},
  \bibinfo {author} {\bibfnamefont {T.}~\bibnamefont {Koishi}}, \bibinfo
  {author} {\bibfnamefont {T.}~\bibnamefont {Ebisuzaki}}, \ and\ \bibinfo
  {author} {\bibfnamefont {X.~C.}\ \bibnamefont {Zeng}},\ }\href@noop {}
  {\bibfield  {journal} {\bibinfo  {journal} {J. Am. Chem. Soc.}\ }\textbf
  {\bibinfo {volume} {135}},\ \bibinfo {pages} {8616} (\bibinfo {year}
  {2013})}\BibitemShut {NoStop}%
\bibitem [{\citenamefont {Oba}\ \emph {et~al.}(2005)\citenamefont {Oba},
  \citenamefont {Okada},\ and\ \citenamefont
  {Maruyama}}]{ObaOkadaMaruyama2005}%
  \BibitemOpen
  \bibfield  {author} {\bibinfo {author} {\bibfnamefont {M.}~\bibnamefont
  {Oba}}, \bibinfo {author} {\bibfnamefont {S.}~\bibnamefont {Okada}}, \ and\
  \bibinfo {author} {\bibfnamefont {S.}~\bibnamefont {Maruyama}},\ }in\
  \href@noop {} {\emph {\bibinfo {booktitle} {29th Fullerene Nanotube General
  Symposium}}}\ (\bibinfo {organization} {Citeseer},\ \bibinfo {year}
  {2005})\BibitemShut {NoStop}%
\bibitem [{\citenamefont {Bhethanabotla}\ and\ \citenamefont
  {Steele}(1987)}]{BhethanabotlaSteele1987}%
  \BibitemOpen
  \bibfield  {author} {\bibinfo {author} {\bibfnamefont {V.~R.}\ \bibnamefont
  {Bhethanabotla}}\ and\ \bibinfo {author} {\bibfnamefont {W.~A.}\ \bibnamefont
  {Steele}},\ }\href@noop {} {\bibfield  {journal} {\bibinfo  {journal}
  {Langmuir}\ }\textbf {\bibinfo {volume} {3}},\ \bibinfo {pages} {581}
  (\bibinfo {year} {1987})}\BibitemShut {NoStop}%
\bibitem [{\citenamefont {Berendsen}\ \emph {et~al.}(1987)\citenamefont
  {Berendsen}, \citenamefont {Grigera},\ and\ \citenamefont
  {Straatsma}}]{BerendsenGrigeraStraatsma1987}%
  \BibitemOpen
  \bibfield  {author} {\bibinfo {author} {\bibfnamefont {H.~J.~C.}\
  \bibnamefont {Berendsen}}, \bibinfo {author} {\bibfnamefont {J.~R.}\
  \bibnamefont {Grigera}}, \ and\ \bibinfo {author} {\bibfnamefont {T.~P.}\
  \bibnamefont {Straatsma}},\ }\href@noop {} {\bibfield  {journal} {\bibinfo
  {journal} {J. Phys. Chem.}\ }\textbf {\bibinfo {volume} {91}},\ \bibinfo
  {pages} {6269} (\bibinfo {year} {1987})}\BibitemShut {NoStop}%
\bibitem [{\citenamefont {Essmann}\ \emph {et~al.}(1995)\citenamefont
  {Essmann}, \citenamefont {Perera}, \citenamefont {Berkowitz}, \citenamefont
  {Darden}, \citenamefont {Lee},\ and\ \citenamefont
  {Pedersen}}]{EssmannPereraBerkowitzDardenLeePedersen1995}%
  \BibitemOpen
  \bibfield  {author} {\bibinfo {author} {\bibfnamefont {U.}~\bibnamefont
  {Essmann}}, \bibinfo {author} {\bibfnamefont {L.}~\bibnamefont {Perera}},
  \bibinfo {author} {\bibfnamefont {M.~L.}\ \bibnamefont {Berkowitz}}, \bibinfo
  {author} {\bibfnamefont {T.}~\bibnamefont {Darden}}, \bibinfo {author}
  {\bibfnamefont {H.}~\bibnamefont {Lee}}, \ and\ \bibinfo {author}
  {\bibfnamefont {L.~G.}\ \bibnamefont {Pedersen}},\ }\href@noop {} {\bibfield
  {journal} {\bibinfo  {journal} {J. Chem. Phys.}\ }\textbf {\bibinfo {volume}
  {103}},\ \bibinfo {pages} {8577} (\bibinfo {year} {1995})}\BibitemShut
  {NoStop}%
\bibitem [{\citenamefont {Miller~III}\ \emph {et~al.}(2002)\citenamefont
  {Miller~III}, \citenamefont {Eleftheriou}, \citenamefont {Pattnaik},
  \citenamefont {Ndirango}, \citenamefont {Newns},\ and\ \citenamefont
  {Martyna}}]{MillerIIIEleftheriouPattnaikNdirangoNewnsMartyna2002}%
  \BibitemOpen
  \bibfield  {author} {\bibinfo {author} {\bibfnamefont {T.~F.}\ \bibnamefont
  {Miller~III}}, \bibinfo {author} {\bibfnamefont {M.}~\bibnamefont
  {Eleftheriou}}, \bibinfo {author} {\bibfnamefont {P.}~\bibnamefont
  {Pattnaik}}, \bibinfo {author} {\bibfnamefont {A.}~\bibnamefont {Ndirango}},
  \bibinfo {author} {\bibfnamefont {D.}~\bibnamefont {Newns}}, \ and\ \bibinfo
  {author} {\bibfnamefont {G.~J.}\ \bibnamefont {Martyna}},\ }\href@noop {}
  {\bibfield  {journal} {\bibinfo  {journal} {J. Chem. Phys.}\ }\textbf
  {\bibinfo {volume} {116}},\ \bibinfo {pages} {8649} (\bibinfo {year}
  {2002})}\BibitemShut {NoStop}%
\bibitem [{\citenamefont {Koishi}\ and\ \citenamefont
  {Takeichi}(2015)}]{KoishiTakeichi2015}%
  \BibitemOpen
  \bibfield  {author} {\bibinfo {author} {\bibfnamefont {T.}~\bibnamefont
  {Koishi}}\ and\ \bibinfo {author} {\bibfnamefont {H.}~\bibnamefont
  {Takeichi}},\ }\href@noop {} {\bibfield  {journal} {\bibinfo  {journal} {Mol.
  Sim.}\ }\textbf {\bibinfo {volume} {41}},\ \bibinfo {pages} {801} (\bibinfo
  {year} {2015})}\BibitemShut {NoStop}%
\bibitem [{\citenamefont {Nos{\'e}}(2001)}]{Nose2001}%
  \BibitemOpen
  \bibfield  {author} {\bibinfo {author} {\bibfnamefont {S.}~\bibnamefont
  {Nos{\'e}}},\ }\href@noop {} {\bibfield  {journal} {\bibinfo  {journal} {J.
  Phys. Soc. Jpn.}\ }\textbf {\bibinfo {volume} {70}},\ \bibinfo {pages} {75}
  (\bibinfo {year} {2001})}\BibitemShut {NoStop}%
\end{thebibliography}

%

\end{document}